\documentclass[twocolumn,trackchanges]{aastex63}
\usepackage{graphicx}
\usepackage{color} 
\received{-}
\revised{-}
\accepted{-}
\submitjournal{ApJ}
\shorttitle{Eclipsing binary distance to the SMC}
\shortauthors{Graczyk et al.}

\begin{document}
\title{A distance determination to the Small Magellanic Cloud with an accuracy of better than 2 percent
         based on late-type eclipsing binary stars}
\correspondingauthor{Dariusz Graczyk}
\email{darek@ncac.torun.pl, darek@astro-udec.cl}

\author[0000-0002-7355-9775]{Dariusz Graczyk}
\affiliation{Centrum Astronomiczne im. Miko{\l}aja Kopernika, Polish Academy of Sciences, Rabia{\'n}ska 8, 87-100, Toru{\'n}, Poland}
\author{Grzegorz Pietrzy{\'n}ski}
\affiliation{Centrum Astronomiczne im. Miko{\l}aja Kopernika, Polish Academy of Sciences, Bartycka 18, 00-716 Warsaw, Poland}
\author{Ian B. Thompson}
\affiliation{Carnegie Observatories, 813 Santa Barbara Street, Pasadena, CA 91101-1292, USA}
\author{Wolfgang Gieren}
\affiliation{Universidad de Concepci{\'o}n, Departamento de Astronomia, Casilla 160-C, Concepci{\'o}n, Chile}
\author{Bart{\l}omiej Zgirski}
\affiliation{Centrum Astronomiczne im. Miko{\l}aja Kopernika, Polish Academy of Sciences, Bartycka 18, 00-716 Warsaw, Poland}
\author{Sandro Villanova}
\affiliation{Universidad de Concepci{\'o}n, Departamento de Astronomia, Casilla 160-C, Concepci{\'o}n, Chile}
\author{Marek G{\'o}rski}
\affiliation{Centrum Astronomiczne im. Miko{\l}aja Kopernika, Polish Academy of Sciences, Bartycka 18, 00-716 Warsaw, Poland}
\author{Piotr Wielg{\'o}rski}
\affiliation{Centrum Astronomiczne im. Miko{\l}aja Kopernika, Polish Academy of Sciences, Bartycka 18, 00-716 Warsaw, Poland}
\author{Paulina Karczmarek}
\affiliation{Universidad de Concepci{\'o}n, Departamento de Astronomia, Casilla 160-C, Concepci{\'o}n, Chile}
\author{Weronika Narloch}
\affiliation{Universidad de Concepci{\'o}n, Departamento de Astronomia, Casilla 160-C, Concepci{\'o}n, Chile}
\author{Bogumi{\l} Pilecki}
\affiliation{Centrum Astronomiczne im. Miko{\l}aja Kopernika, Polish Academy of Sciences, Bartycka 18, 00-716 Warsaw, Poland}
\author{Monica Taormina}
\affiliation{Centrum Astronomiczne im. Miko{\l}aja Kopernika, Polish Academy of Sciences, Bartycka 18, 00-716 Warsaw, Poland}
\author{Rados{\l}aw Smolec}
\affiliation{Centrum Astronomiczne im. Miko{\l}aja Kopernika, Polish Academy of Sciences, Bartycka 18, 00-716 Warsaw, Poland}
\author{Ksenia Suchomska}
\affiliation{Centrum Astronomiczne im. Miko{\l}aja Kopernika, Polish Academy of Sciences, Bartycka 18, 00-716 Warsaw, Poland}
\author{Alexandre Gallenne}
\affiliation{Centrum Astronomiczne im. Miko{\l}aja Kopernika, Polish Academy of Sciences, Bartycka 18, 00-716 Warsaw, Poland}
\affiliation{Departamento de Astronom{\'i}a, Universidad de Concepci{\'o}n, Casilla160-C, Concepci{\'o}n, Chile}
\affiliation{Universit\'e C$\hat{\rm o}$te d'Azur, Observatoire de la C$\hat{\rm o}$te d'Azur, CNRS, Laboratoire Lagrange, Nice, France}
\affiliation{Unidad Mixta Internacional Franco-Chilena de Astronom{\'i}a (CNRS UMI 3386), Departamento de Astronom{\'i}a, Universidad de Chile, Camino El Observatorio 1515, Las Condes, Santiago, Chile}
\author{Nicolas Nardetto}
\affiliation{Universit\'e C$\hat{\rm o}$te d'Azur, Observatoire de la C$\hat{\rm o}$te d'Azur, CNRS, Laboratoire Lagrange, Nice, France}
\author{Jesper Storm}
\affiliation{Leibniz-Institut f\"{u}r Astrophysik Potsdam, An der Sternwarte 16, 14482 Potsdam, Germany}
\author{Rolf-Peter Kudritzki}
\affiliation{Institute for Astronomy, University of Hawaii at Manoa, 2680 Woodlawn Drive, Honolulu, HI 96822-1897, USA}
\affiliation{LMU M{\"u}nchen, Universit{\"a}tssternwarte, Scheinerstr. 1, D-81679 M{\"u}nchen, Germany}
\author{Miko{\l}aj Ka{\l}uszy{\'n}ski}
\affiliation{Centrum Astronomiczne im. Miko{\l}aja Kopernika, Polish Academy of Sciences, Bartycka 18, 00-716 Warsaw, Poland}
\author{Wojciech Pych}
\affiliation{Centrum Astronomiczne im. Miko{\l}aja Kopernika, Polish Academy of Sciences, Bartycka 18, 00-716 Warsaw, Poland}
\begin{abstract}

We present a new study of late-type eclipsing binary stars in the Small Magellanic Cloud (SMC) undertaken with the aim of improving the distance determination to this important galaxy. A sample of 10 new detached, double-lined eclipsing binaries indentified from the OGLE variable star catalogues and consisting of F- and G-type giant components has been analysed. The absolute physical parameters of the individual components have been measured with a typical accuracy of better than 3\%.  All but one of the systems consist of young and intermediate population stars with masses in the range of 1.4 to 3.8~$M_\sun$.  

This new sample has been combined with five SMC eclipsing binaries previously published by our team. Distances to the binary systems were calculated using a surface brightness - color calibration. The targets form an elongated structure, highly inclined to the plane of the sky. The distance difference between the nearest and most-distant system amounts to 10 kpc with the line of sight depth reaching 7 kpc. We find tentative evidence of the existence of a spherical stellar sub-structure (core) in the SMC coinciding with its stellar center, containing about 40\% of the young and intermediate age stars in the galaxy. The radial extension of this sub-structure is $\sim 1.5$ kpc. We derive a distance to the SMC center of $D_{\rm SMC}=62.44 \pm 0.47 $(stat.)$ \pm 0.81$ (syst.) kpc corresponding to a distance modulus $(m-M)_{\rm SMC}=18.977 \pm 0.016\pm 0.028$ mag, representing an accuracy of better than 2\%. \\
\end{abstract} 

\section{Introduction}
The Small Magellanic Cloud (SMC) is one of the nearest satellite galaxies of the Milky Way.  Together with the Large Magellanic Cloud (LMC) this irregular dwarf forms a pair of interacting galaxies, and growing evidence indicates that the SMC is experiencing a tidal disruption which will  eventually lead to its complete disintegration. These two galaxies are connected by the Magellanic Bridge containing gas and young and intermediate aged stars. The SMC is probably on a very eccentric orbit around the LMC and the galaxies collided in a past encounter $\sim150$ Myr \citep[e.g.][]{ziv18}. The collision/interaction left the LMC mostly intact but the less massive  SMC has been profoundly  disturbed. Of principal importance is a knowledge of the precise distances to both galaxies, which contribute to an understanding of their orbits, the history of mutual interactions, and their future fates.

In recent years there has been additional interest in the precise determination of the distance to the SMC  because  this galaxy is key in quantifying the population effects of a number of distance indicators. Although the distance to the LMC remains the prime calibrator of the extragalactic distance scale \citep[e.g.][]{rie19,fre19}, the low mean metallicity (smaller by a factor of $\sim3$ than the LMC) together with its richness in variable stars make the SMC an important secondary calibrator.
In practice most of distance determination methods can be applied to both Magellanic Clouds, permitting a careful determinations of systematic effects. Knowledge of the SMC distance in particular allows for a  determination of population/metallicity effects in Period-Luminosity relations of classical Cepheids  \citep[e.g.][]{wie17,gie18}, a measurement of the projection factor of Cepheids \citep{gal17}, a determination of the brightness of the tip of the red giant branch (TRGB) \citep[e.g.][]{gor18,gro19,fre20}, a calibration of the H$\alpha$ surface brightness-radius relation for planetary nebulae \citep{fre16}, a measurement of the empirical upper luminosity boundary of cool supergiants \citep{dav18}, and a calibration of  J-Branch Asymptotic Giant Branch star luminosities \citep[e.g.][]{mad20}.

The complex geometry and the large extension in the line of sight of the SMC makes an accurate determination of its distance difficult. For example, the distances of classical Cepheids in the SMC vary by about 15 kpc \citep{gie18}. Even defining the center of this galaxy is difficult \citep[see e.g.][]{deG15}. Conflicting distances and large accompanying errors were one of the main motivations for using eclipsing binaries to determine the distance to the SMC. \cite{bel91} presented the first attempt to precisely determine the physical parameters and distance of an extragalactic eclipsing binary star. For many years eclipsing binary distances to the SMC were based on systems containing early-type stars, very often in a semidetached configuration \citep[e.g.][]{pri98,har03,hil05,nor10}. Detached eclipsing binaries offer a simpler and less model dependent geometry, and binaries containing late-type stars have the advantage in that accurate empirical surface brightness - color relations can be used to derive distances with an accuracy of  $\sim$$2\%$ or better \citep[e.g.][]{tho01,pie09,gra12}. 

This present work is a culmination of an 18-year long program to identify  and study very rare detached systems containing photometrically stable late-type giant stars in the SMC. Here we report on the analysis and distance determination to 10 new  eclipsing binaries in the SMC. This study is an extension and improvement to our previous study where only 5 stars were used to determine an SMC distance \citep[hereafter G14]{gra14}, and is a complement to our work on an eclipsing binary distance to the LMC \citep{pie19}.

\section{Observations and Data Reduction}

\begin{deluxetable*}{@{}llcclcCCCc@{}}
\tabletypesize{\scriptsize}
\tablecaption{The target stars \label{tbl:1}}
\tablewidth{0pt}
\tablehead{
\colhead{OGLE ID} & \colhead{MACHO ID} & \colhead{RA} & \colhead{Dec} & \colhead{$V$} & \colhead{$I$} & \colhead{$J$} & \colhead{$K$} & \colhead{$P_{\rm obs}$}  & \colhead{$T_0$} \\ 
\colhead{} & \colhead{}& \colhead{h:m:s} & \colhead{deg:m:s} & \colhead{mag} & \colhead{mag} & \colhead{mag} &\colhead{mag}& \colhead{d} & \colhead{HJD}
}
\startdata
SMC-ECL-0019 & $-$		    &00:22:55.52 &$-$73:53:16.5 &  17.921(10) & 16.907(6) & 16.234(33) & 15.660(21) &143.9982&2456673.421\\
SMC-ECL-0439 & $-$ 		    &00:40:21.45 &$-$73:27:19.5 & 18.061(10) & 17.117(6)  & 16.501(24) & 15.970(24) &279.3848&2455088.926 \\
SMC-ECL-0727 & 212.15619.175        &00:44:12.10 & $-$73:17:42.0 & 18.109(10) & 17.022(6) & 16.190(79) & 15.671(16) & \!\!\!316.744&$\!\!\!$2455133.80 \\
SMC-ECL-0970 & 208.15743.89         &00:46:00.29 & $-$72:39:01.0 &  17.926(9)&  16.810(6) & 16.044(22) & 15.367(21) &191.6426&2455186.538\\
SMC-ECL-1492 & 212.15848.1239       &00:48:41.92 & $-$73:16:18.0 &  17.707(9)& 16.626(6)  & 15.941(31) & 15.219(23) &\;\;73.7567&2452129.701 \\
SMC-ECL-1859 & 212.15957.68         &00:50:20.57 &$-$73:35:37.7  &17.685(9) &16.760(6)    & 16.166(19) & 15.587(16) &\;\;75.5805 &2456568.143 \\
SMC-ECL-2761 & 207.16198.118        &00:53:31.46 & $-$72:42:58.0 & 17.296(8)& 16.460(5)   & 15.930(75) & 15.406(29) &150.4410 &2452025.820\\
SMC-ECL-2876 & $-$		    &00:53:58.59 & $-$71:37:21.1 &  17.486(6) & 16.589(5) & 16.017(13) & 15.529(11) &120.9087 &2456682.792 \\
SMC-ECL-3529 & 211.16418.53         &00:57:05.35 & $-$73:15:10.8 &  17.116(6)& 16.291(5)  & 15.797(23) & 15.333(15) &234.4648 &2455039.612 \\
SMC-ECL-3678 & 211.16475.2	    &00:57:49.59 & $-$73:15:46.5 & 15.558(5) & 14.826(3)  & 14.282(36) & 13.800(20) &187.9557 &2452023.894 
\enddata
\tablecomments{Columns give coordinates, observed magnitudes, colors, orbital periods and epochs of the primary minimum.  Identification numbers are from the OGLE-III catalogue of variable stars \citep{paw13,paw16}. Observed SOFI $J$, $K$ magnitudes are expressed in the 2MASS photometric system.}
\end{deluxetable*}

\subsection{Selected Targets}
We selected our targets from  catalogues of eclipsing binary stars in the SMC \citep{wyr04,paw13,paw16} which are based on OGLE-II, -III and -IV data \citep{uda97,uda03,uda15}. We adopted a lower magnitude cut of 18.2 mag in the $V$-band. We chose systems with orbital periods longer than 45 days and having red colors $(V\!-\!I)>0.5$ mag. A total of 10 clearly detached systems were left in the sample. The candidates were confirmed as double-lined binaries with spectroscopic observations.  Basic data for the targets are given in Table~\ref{tbl:1}.

\label{obs}
 
 \subsection{Photometric Data}
 \label{obs:phot}
The optical $V$- and $I$-band light curves were collected from  published OGLE data \citep{paw13,paw16}. We augmented these data with the MACHO $B$- and $R$-band light curves \citep{alc99,fac07}. The light curves were cleaned of outliers, especially the MACHO light curves, and were detrended.

Near-infrared $J$- and $K$-band photometry was collected with the ESO NTT telescope on La Silla, equipped with the SOFI camera \citep{moo98}. The setup of the instrument, reduction, and calibration of the data onto the UKIRT system are described in \cite{pie09}.  We collected at least two epochs of infrared photometry for each of our targets outside of eclipses.
In order to validate our data and quantify the zero-point uncertainty of $K$-band magnitudes we added near-infrared $J$- and $K$-band photometry from the Vista Magellanic Cloud (VMC) survey \citep{cio11}, the Infrared Survey Facility (IRSF) \citep{kat07}, and the Two Micron  All Sky Survey (2MASS) 6X Point Source Catalog \citep{cut12}. The photometry was later converted onto the 2MASS system using the transformation equations given by \cite{car01}.
 
\subsection{Spectroscopic Data} 
High-resolution echelle spectra were collected with the MIKE spectrograph on the Magellan Clay 6.5-m telescope at Las Campanas \citep{ber03}, the HARPS spectrograph on the 3.6-m telescope at La Silla \citep{may03}, and the UVES spectrograph on the 8.4-m VLT Unit 2  telescope at Paranal \citep{dek00}. The MIKE data were obtained with  a 5$\times$0$\,\farcs$7 slit, yielding  a resolution of about 40,000. In the case of HARPS, we used the EGGS mode, giving a resolution of about R $\sim$80,000. UVES was operated in the high-resolution mode with R $\sim$80,000. Integration times were typically 1 hour for MIKE and 30 minutes for UVES, depending on observing conditions and the magnitude of the star. The HARPS spectra were reduced by the on-line pipeline, the UVES spectra were reduced using \verb"UVES Workflow version 5.8.2" pipeline provided by ESO \citep{fre13}, and MIKE spectra were reduced using Daniel Kelson's pipeline available at the Carnegie Observatories software repository.
 
\subsection{Radial Velocities}
The radial velocities were extracted from reduced 1-dimensional spectra using the RaVeSpAn code \citep{pil17}. We employed the Broadening Function method \citep{ruc92,ruc99}. The templates were interpolated from grids of theoretical stellar spectra \citep{col05}. Progress in the analysis of individual systems resulted in improvements in the surface temperatures and gravities which were then used to recalculate the templates and to re-derive the radial velocities. We assumed an SMC metallicity of [Fe/H]$ = -0.8$ for all templates \citep[e.g.][]{pia12}. The final radial velocities are presented in Table~\ref{tbl:spec}. The errors of radial velocities are measured from the broadening function profiles and the typical precision of an individual measurement is about 200 m s$^{-1}$.

\begin{deluxetable}{@{}l@{}r@{}c@{}c@{}c@{}c@{}l}
\tabletypesize{\scriptsize}
\tablecaption{The radial velocity
 measurements \label{tbl:spec}}
\tablewidth{0pt}
\tablehead{
\colhead{OGLE ID} & \colhead{HJD$-$} & \colhead{RV1} & \colhead{Err1} & \colhead{RV2} & \colhead{Err2} &\colhead{Instr.}  \\
\colhead{$\!\!\!\!$SMC-ECL-} & \colhead{$-$2450000} & \colhead{km s$^{-1}$} & \colhead{km s$^{-1}$} &  \colhead{km s$^{-1}$} & \colhead{km s$^{-1}$} & \colhead{}
}
\startdata
0019  & 6579.64339 & 99.41 & 0.20& 148.94 & 0.20 & MIKE \\ 
0019  & 7359.70725 & 152.61 & 0.20 & 91.32 & 0.20 & UVES \\ 
0019  & 7359.72874 & 151.98 & 0.34 & 91.44 & 0.30 & UVES \\ 
0019  & 7642.56023 & 152.64 & 0.18 & 91.54 & 0.18 & UVES\\ 
0019  & 7658.68864 & 147.39 & 0.18 & 96.57 & 0.18 & UVES 
\enddata
\tablecomments{This table is available entirety in electronic format in the online journal. A portion is shown here for guidance regarding its form and content.}
\end{deluxetable}

\section{Modeling the systems}


\begin{deluxetable}{@{}l@{}C@{}C@{}C@{}C@{}C@{}}
\tabletypesize{\scriptsize}
\tablecaption{The spectroscopic light ratios at 5500 \AA$\,$ (V-band) or/and at 6400~\AA$\,$ (R$_C$ band) \label{tbl:spec2}}
\tablewidth{0pt}
\tablehead{
\colhead{OGLE ID} & \twocolhead{Line intensity } & \colhead{Correction} & \twocolhead{Light ratio} \\
\colhead{SMC-} & \colhead{$I_2/I_1(\rm{V})$} & \colhead{$I_2/I_1(\rm{R}_C)$} & \colhead{$k_{21}$} & \colhead{$L_2/L_1(\rm{V})$} &\colhead{$L_2/L_1(\rm{R}_C)$}
}
\startdata
ECL-0019 &-&1.15\pm0.06&$0.97 \pm 0.02$&-&1.12\pm0.06\\
ECL-0439 & $1.04 \pm 0.13$ &-& $1.00 \pm 0.02$&$1.04 \pm 0.13$&-\\
ECL-0727 & - &1.73\pm0.05& $0.75 \pm 0.04$ &-&1.30\pm0.08\\
ECL-0970 &-&1.34\pm0.04$&0.95 \pm 0.03$&-&1.27\pm0.06\\
ECL-1492&$2.33\pm0.17$&-&$0.90 \pm 0.05$&$2.10 \pm 0.20$&-\\
ECL-1859&-&1.12\pm0.04&$ 0.72 \pm 0.04$&-&0.81 \pm 0.05\\
ECL-2761& $1.01 \pm 0.04$ &-& $0.96 \pm 0.02$ &$0.97\pm0.04$&-\\
ECL-2876& $0.91\pm0.05$&-&$0.81 \pm 0.03$&$0.74 \pm0.05$&-\\
ECL-3529&-$&2.86\pm0.19&$0.81\pm0.03$&&2.32\pm0.18\\
ECL-3678& $1.33 \pm 0.05$ &0.88\pm0.03& $0.51 \pm 0.04$&$0.45\pm0.04$&0.68\pm0.06
\enddata
\tablecomments{Line intensity - the relative strengths of the absorption lines of the secondary with respect to those of the primary.}
\end{deluxetable}

\begin{deluxetable*}{@{}lcccccccc}
\tabletypesize{\scriptsize}
\tablecaption{Atmospheric parameters\label{tbl:atmo}}
\tablewidth{0pt}
\tablehead{
\colhead{OGLE ID} &\colhead{}&\colhead{$T_{\rm a,eff}$}& \colhead{$[$Fe/H$]$}&\colhead{$v_{mt}$}&\colhead{$T_{\rm c,eff}$}&\colhead{$[$Fe/H$]_{\rm C}$}\\
& &\colhead{K}  & \colhead{dex} &\colhead{km $s^{-1}$} &\colhead{K}&\colhead{dex}
}
\startdata
SMC-ECL-0019 &p&5310&$-$0.53&4.3 &5130&$-$0.60\\
			  &s 	&  $-$ &  $-$  & 4.3&5035&$-$\\   
SMC-ECL-0439 &p	& 5060   &  $-$1.46 & 4.3 &5235&$-$1.34\\
                            &s& 5275   & $-$1.16 & 4.3 &5250&$-$1.17\\
SMC-ECL-0727 &p	&$-$    &  $-$    &4.4&5235&$-$\\
                            &s& 4940    &  $-$0.86    &3.8&4800&$-$0.93\\
SMC-ECL-0970 &p	& 4910    & $-$1.02   &3.8 &4825&$-$1.06\\
                            &s&4790 &   $-$0.59  &3.8&4760&$-$0.60\\
SMC-ECL-1492  &p  &   5580 &  $-$0.30   &4.0&4985&$-$0.69\\
                            &s& 4740& $-$1.09&4.0&4785&$-$1.07\\                           
SMC-ECL-1859  &p	& 5850  & $-0.02$ & 4.9 &5535&$-0.23$\\
                            &s& $-$   & $-$  &4.0 &4935&$-$\\
SMC-ECL-2761  &p	&  6030    &  $-$0.66    &5.2&5625&$-$0.92\\
                            &s& 5195   &  $-$0.79   &5.1&5575&$-$0.54\\
SMC-ECL-2876  &p	&  5740  &  $-$0.34   &5.0&5550&$-$0.46\\
                            &s&4970  &  $-$0.52 &4.2 &5140&$-$0.40\\
SMC-ECL-3529  &p 	&  5835   & $-$0.93   &5.3&5815&$-$0.93\\
                            &s& 5110  &  $-$1.05 & 4.7 &5445&$-$0.93\\
SMC-ECL-3678   &p & 6790    &  $-$0.75    &6.0&6740&$-$0.84\\
                            &s& 5040    &  $-$0.68    &4.7&4915&$-0.72$
 \enddata
 \tablecomments{$T_{\rm a,eff}$ is the spectroscopic temperature (Sec.~\ref{sec:atmo}),[Fe/H] is the metallicity derived from decomposed spectra, $v_{mt}$ is the macroturbulent velocity,  $T_{\rm c,eff}$ is the color temperature (Sec.~\ref{sec:temp}), [Fe/H]$_C$ is the metallicity corrected for the temperature-metallicity correlation (Sec.~\ref{sec:atmo}).}
 \end{deluxetable*}
 
\subsection{Spectroscopic Light Ratio}
\label{sec:spec}
The line intensity ratio is equal to the ratio of the equivalent widths of absorption lines in the individual components. The ratios of the equivalent widths of absorption lines were  derived for the $V$- and $R_C$-bands from the broadening function profiles  over the wavelength regions   ${\rm 5000\,\AA-6000\,\AA}$ ($V$-band) and ${\rm 5950\,\AA-6850\,\AA}$ ($R_C$-band). The $V$-band ratios are based on MIKE spectra and the $R_C$-band ratios are based on UVES-RED spectra in those cases when we could not calculate a $V$-band ratio of sufficient quality.  We averaged the ratios from all available spectra for any one star in our sample.  The line intensity ratios $I_2/I_1$ are given in Table~\ref{tbl:spec2}. To convert $I_2/I_1$  into light ratios we calculated the corrections $k_{21}$ following \cite{gra18}. The final spectroscopic light ratios corresponding to the true $V$-band or $R_C$-band light ratios were computed simply as the product $k_{21}\cdot I_2/I_1$, and these are also  given in Table~\ref{tbl:spec2}. 

\subsection{Spectral Disentangling and Atmospheric Parameters Analysis}
\label{sec:atmo}
We used the method outlined by \cite{gon06} to disentangle the individual spectra of the binary components. The method is an iterative procedure in which the spectrum of one component is alternately used to calculate the spectrum of the other component. The method works in the real domain. For the renormalization of the disentangled spectra we follow the methodology described in G14. The disentangling was done with the RaVeSpAn code. We used the MIKE and UVES spectra to obtain the disentangled spectra.

Atmospheric parameters and the iron content were obtained from the equivalent widths (EWs) of the iron spectral lines in the wavelength range  5800\,${\rm \AA}$ to 6800\,${\rm \AA}$. See \cite{mar08} for a more detailed explanation of the method  used to measure the EWs and \cite{vil10} for a description of  the line list that was used.  We adopted $\log{\epsilon}(Fe)=7.50$ as the solar iron abundance. The local thermodynamic equilibrium  program MOOG \citep{sne73} was used for the abundance analysis together with Kurucz atmospheric models \citep{kur70}. For more details of the derivation of atmospheric parameters see \cite{gra18}. In a few cases one of the components was significantly weaker than the companion and we could not obtain a decomposed spectrum with sufficient quality to derive atmospheric parameters. Those cases, however, do not  significantly affect the derived individual physical parameters and distances -- see Sec.~\ref{red} and~\ref{dist:ind}.

Some degeneracy exists between the spectroscopic temperature T$_{\rm a,eff}$ and [Fe/H], especially for lower S/N spectra. We measured a mean metallicity - temperature shift of 0.07 dex per 100 K. We used this value to adjust all measured [Fe/H] using the differences between  $T_{\rm a,eff}$ and the final temperatures from our photometric analysis (see Sec.~\ref{fin}). The corrected metallicities [Fe/H]$_{\rm C}$ are reported in the final column of Table~\ref{tbl:atmo}.  The macroturbulent velocity was taken as the mean of values calculated with equations (2-4) given in \cite{gra18} and which are taken from \cite{hek07,tak08,mas08}, respectively.

\begin{deluxetable}{@{}lcccc@{}}
\tabletypesize{\scriptsize}
\tablecaption{Color Excess E($B\!-\!V$) \label{tbl-5}}
\tablewidth{0pt}
\tablehead{\colhead{OGLE ID} &\colhead{Na I D1} &\colhead{Haschke et al.} & \colhead{Atmos.} &\colhead{Adopted}\\
\colhead{} & \colhead{}  & \colhead{$+\,0.027$}  & \colhead{} & \colhead{} \\ 
\colhead{} &\colhead{(mag)}  &\colhead{(mag)} & \colhead{(mag)} &\colhead{(mag)}
}
\startdata
SMC-ECL-0019 & 0.056 & 0.037 & 0.130 & $0.074\pm0.030$ \\
SMC-ECL-0439 & 0.071 & 0.052 & 0.041 & $0.055\pm0.015$ \\
SMC-ECL-0727 & 0.111 & 0.080 & 0.157 & $0.116\pm0.030$ \\
SMC-ECL-0970 & 0.078 & 0.057& 0.102 & $0.079\pm0.018$ \\
SMC-ECL-1492 & 0.073 & 0.071 & 0.063 & $0.069\pm0.012$ \\
SMC-ECL-1859 & 0.069 & 0.049 & - & $0.059\pm0.013$ \\
SMC-ECL-2761 & 0.083 & 0.072 & 0.093 & $0.083\pm0.013$ \\
SMC-ECL-2876 & 0.050 & 0.057& 0.055 & $0.054\pm0.012$ \\
SMC-ECL-3529 & 0.072 & 0.047 & 0.003& $0.041\pm0.029$ \\
SMC-ECL-3678 & 0.059 & 0.047 & 0.096 & $0.067\pm0.021$ 
\enddata
\end{deluxetable}
 
 \subsection{Interstellar Reddening}
 \label{red}
The interstellar extinction in the direction to each of our target stars was derived in three ways following G14. First we used a calibration between the equivalent width of the interstellar  Na I D1 absorption line (5890.0 {\rm \AA}) and the reddening \citep{mun97}. The calibration works best for relatively small values of reddening, E($B\!-\!V$)$\,<0.4$ mag. The method uses an empirical relation between gas and dust content calibrated for our Galaxy. We expect only slight deviations from this relation for the SMC since on average about 60\% of the total reddening comes from foreground Galactic extinction. As a result, any possible systematic offset is small and is included in the statistical error of the final adopted reddening. We calculated the reddening separately for the Galactic absorption component(s) and the SMC absorption component(s) of the Na I D1 line. We then add these together and the second column of Table~\ref{tbl-5} gives the final values for the reddening derived in this way. 

The second method is based on the Magellanic Cloud reddening maps published by \cite{has11}. Although new reddening maps in both Magellanic Clouds have been published recently \citep{gor20,sko20,nat20} in order to keep the same systematics as in our previous works \citep[G14,][]{pie19} we  only used maps from Haschke et al. Inclusion of any of the new reddening maps does not  significantly change results of our paper. For example,  inclusion of maps by \cite{sko20} would change our final reddening determinations by less $-0.001$ mag on average, and  inclusion of maps by \cite{gor20} would change the reddenings by $+0.007$ mag on average. We added $\Delta$E($B\!-\!V$)$\,=0.027$ mag to reddening estimates from the \cite{has11} maps to account for the mean foreground Galactic reddening in the direction of the SMC. The third column of Table~\ref{tbl-5} gives the resulting reddening to each eclipsing binary. 

\begin{figure*}
\begin{minipage}[th]{0.5\linewidth}
\includegraphics[angle=0,scale=.47]{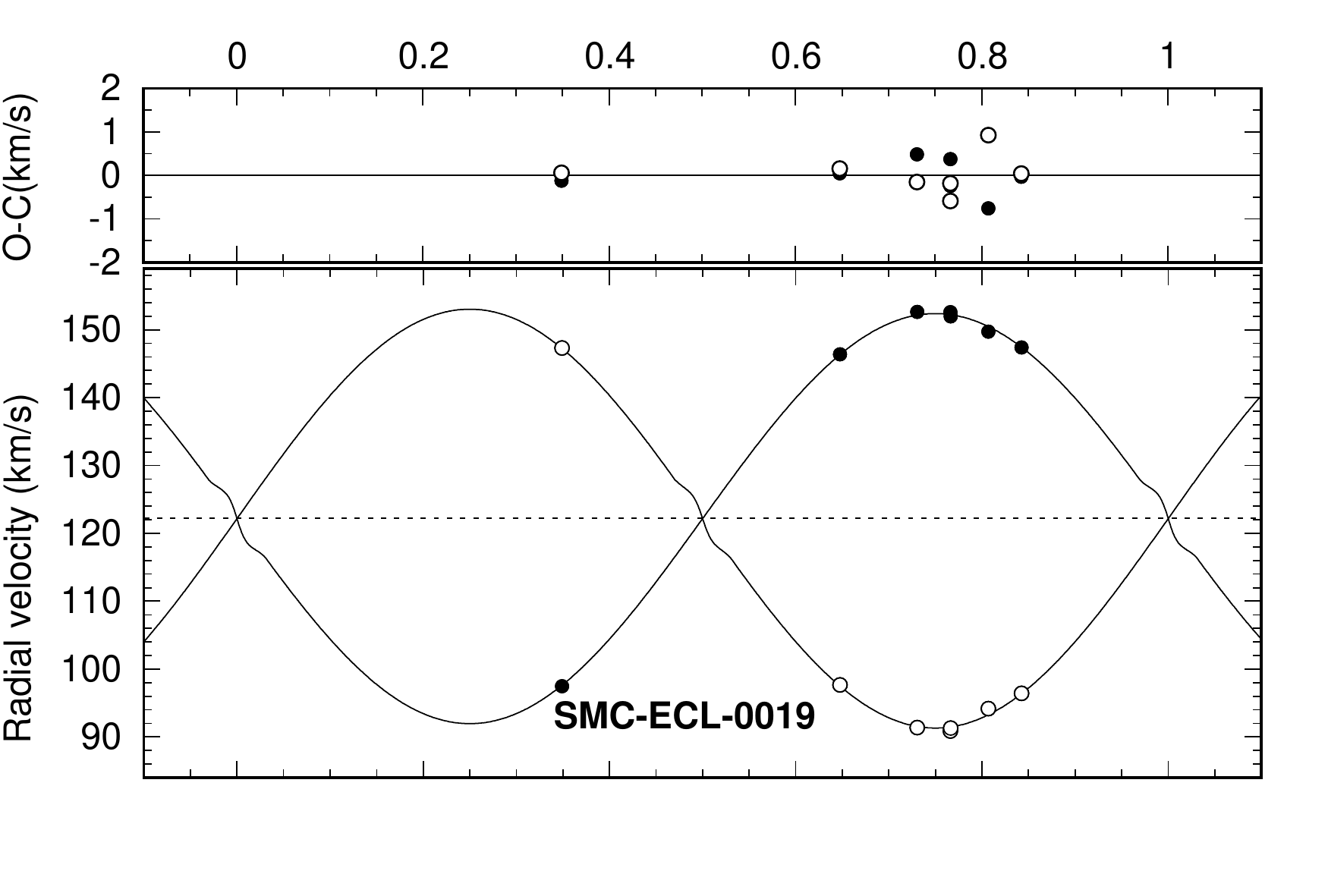} \vspace{-1.0cm}
\mbox{}\\ 
\includegraphics[angle=0,scale=.47]{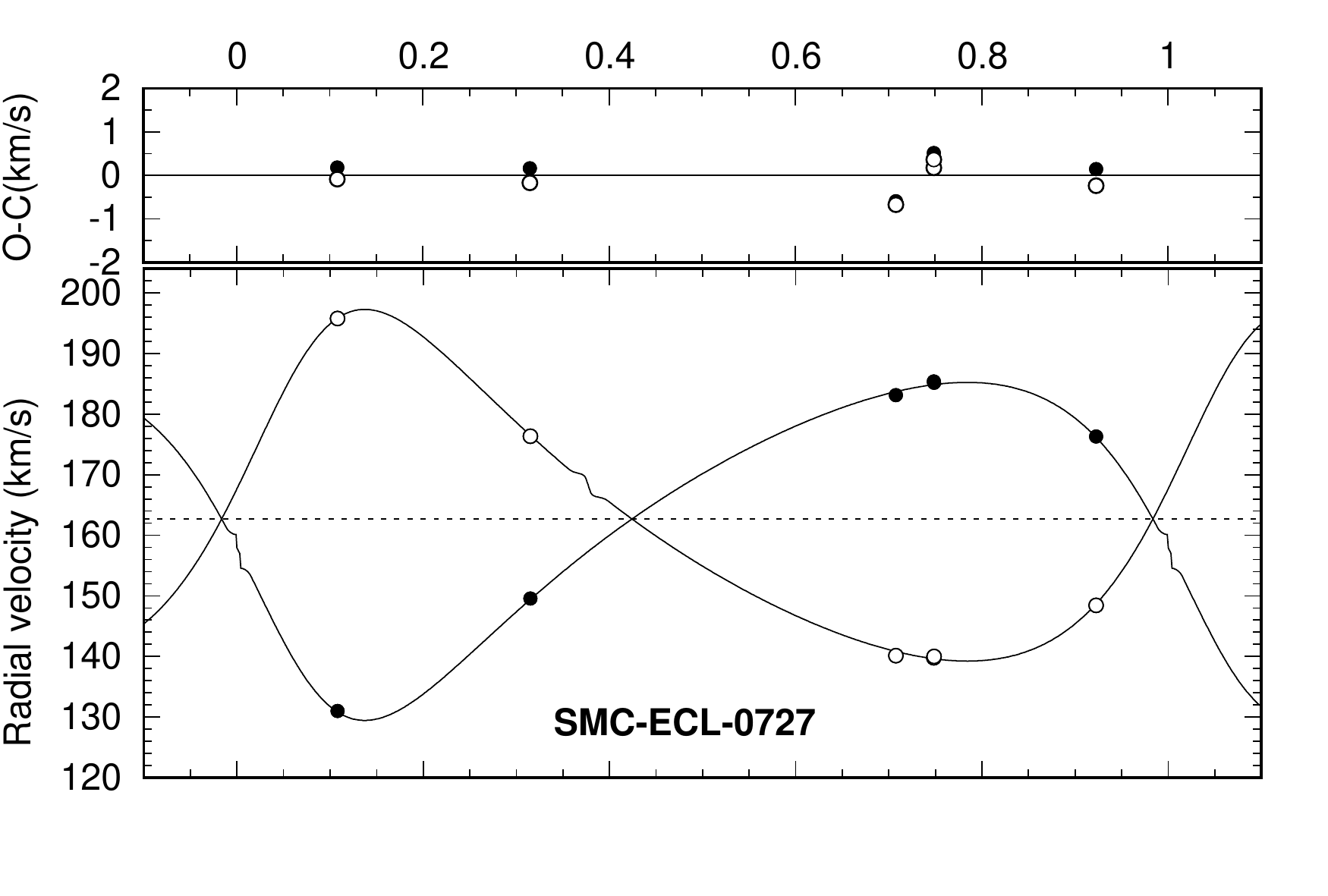}\vspace{-1.0cm}
\mbox{}
\includegraphics[angle=0,scale=.47]{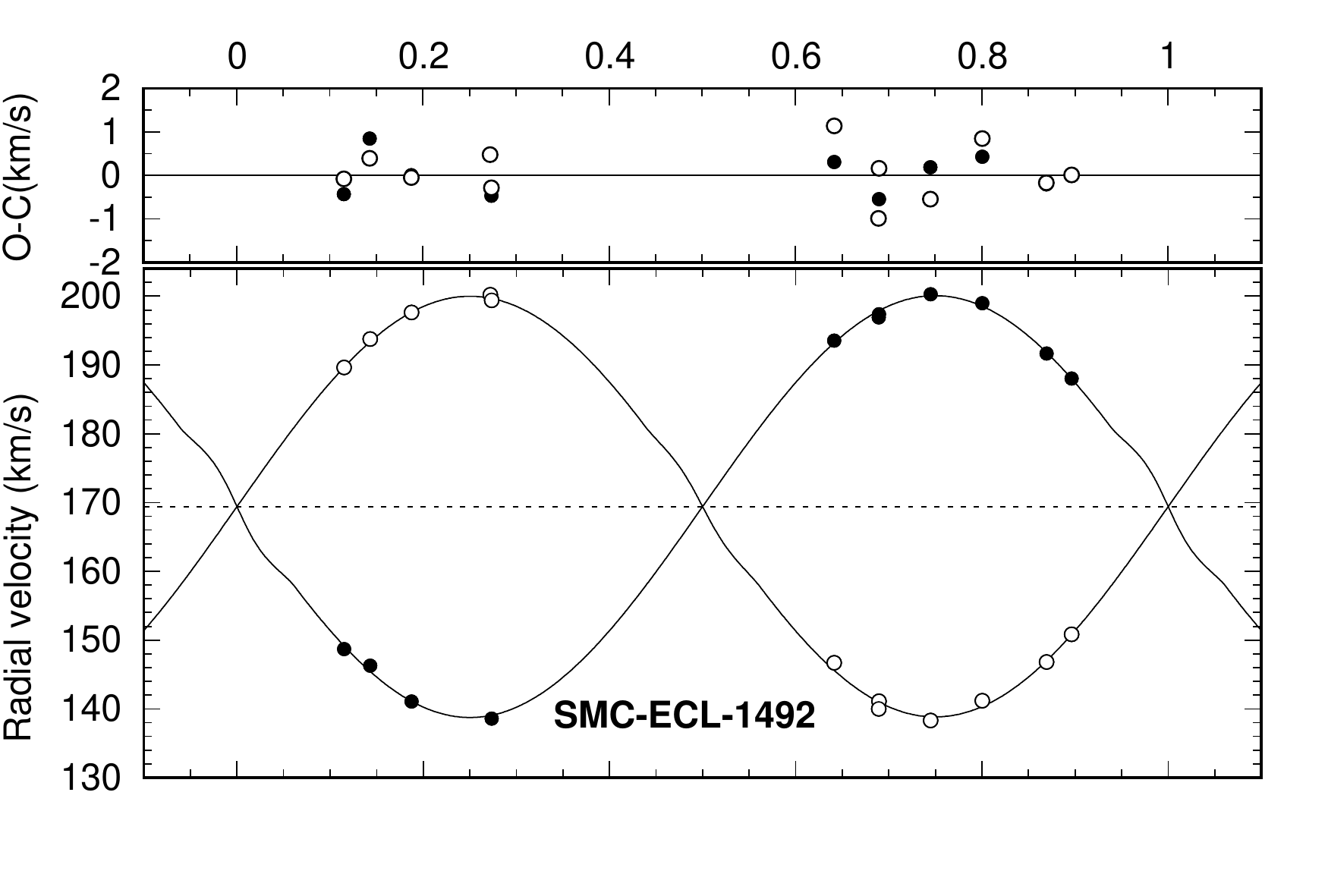}\vspace{-1.0cm}
\mbox{}
\includegraphics[angle=0,scale=.47]{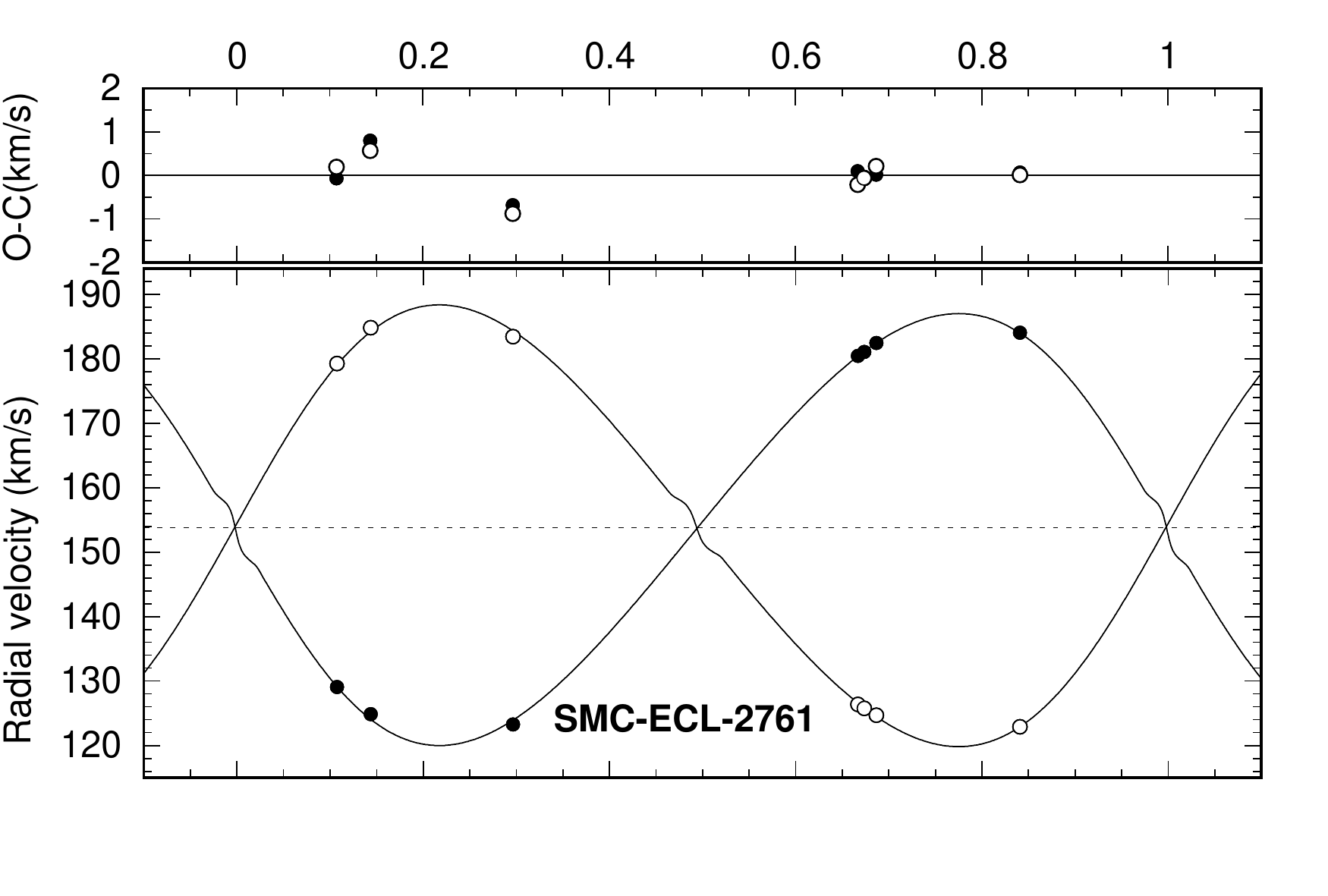} \vspace{-1.0cm}
\mbox{}\\ 
\includegraphics[angle=0,scale=.47]{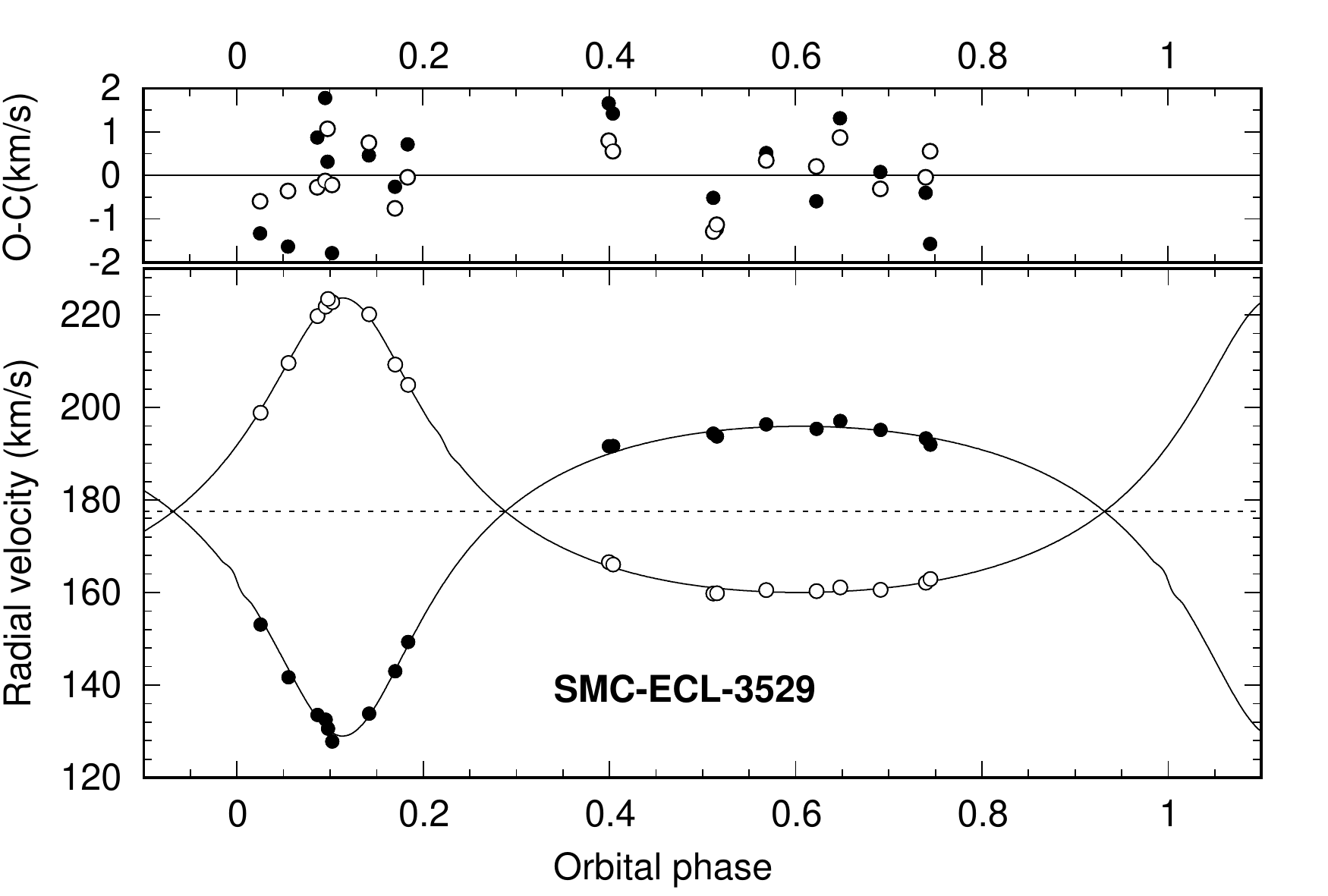}
\end{minipage}\hfill 
\begin{minipage}[th]{0.5\linewidth}
\includegraphics[angle=0,scale=.47]{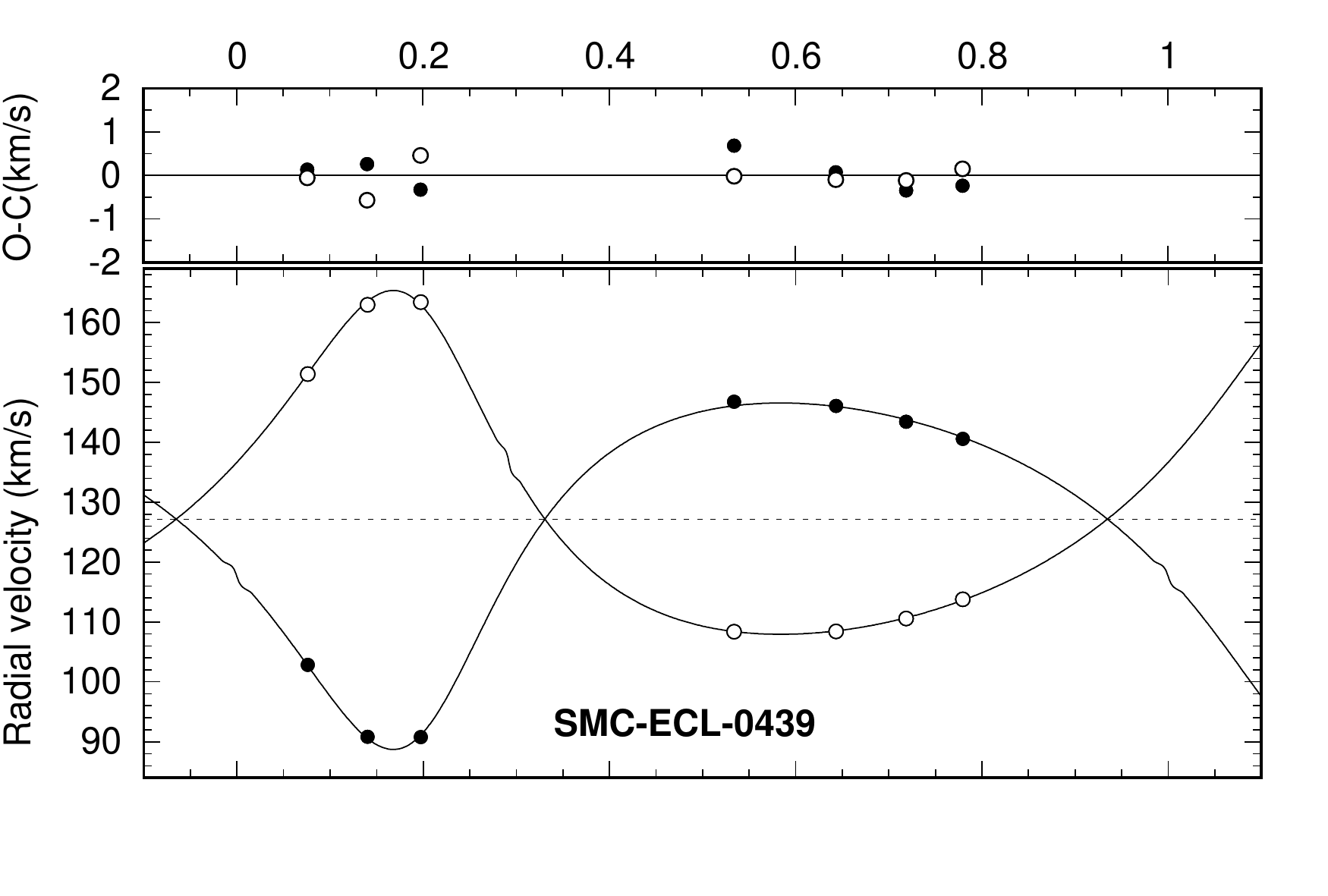}\vspace{-1.0cm}
\mbox{}
\includegraphics[angle=0,scale=.47]{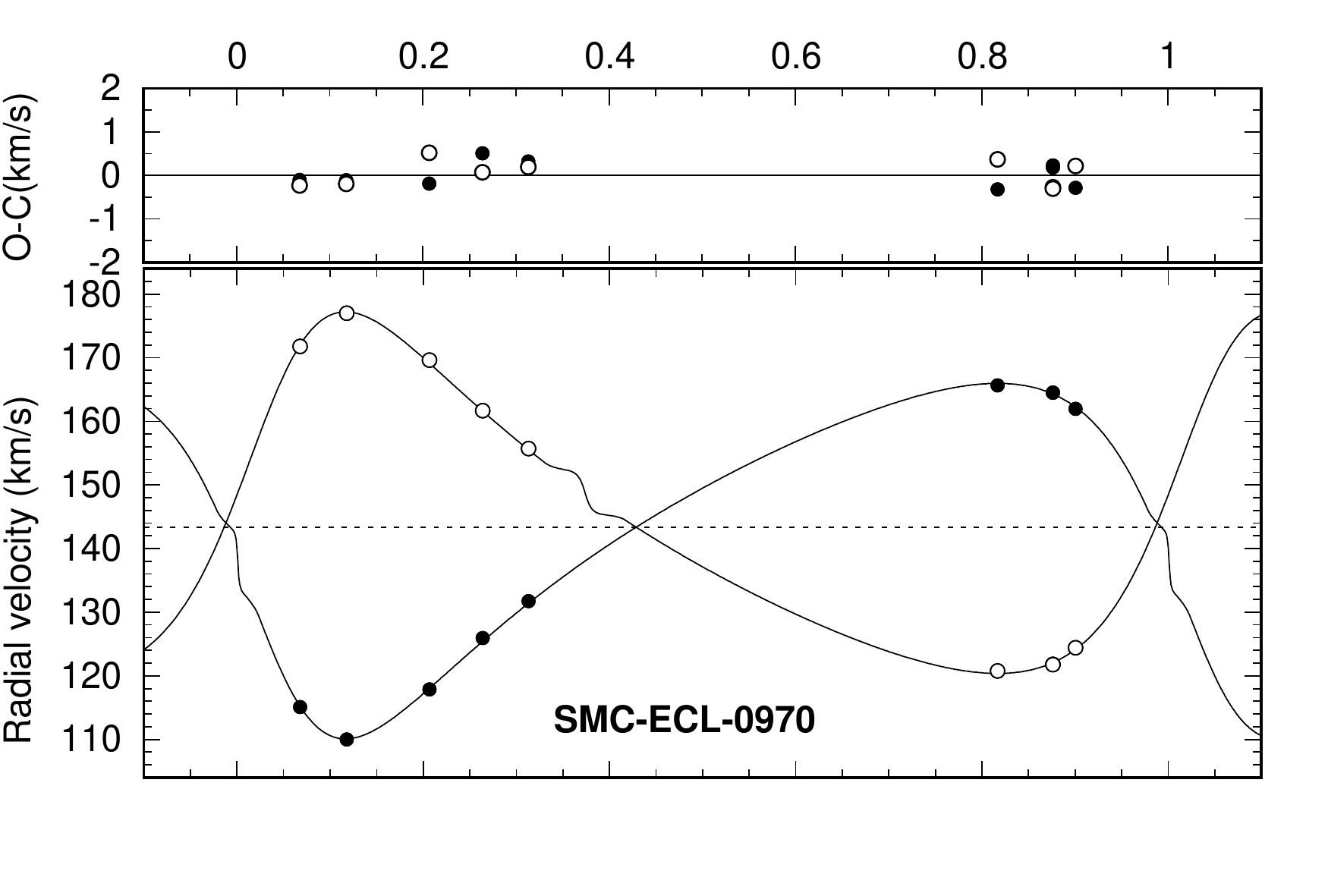}\vspace{-1.0cm}
\mbox{}
\includegraphics[angle=0,scale=.47]{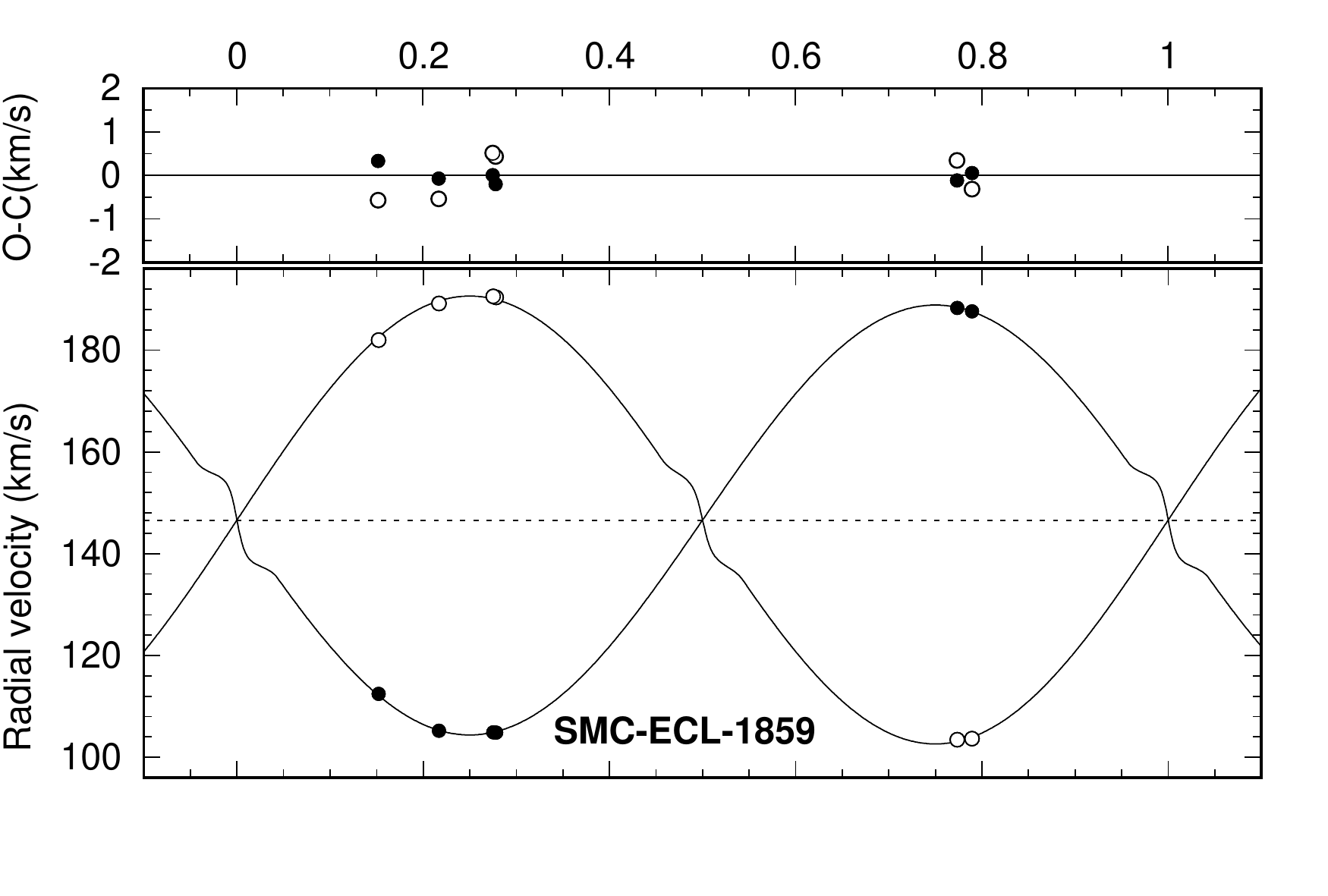}\vspace{-1.0cm}
\mbox{}
\includegraphics[angle=0,scale=.47]{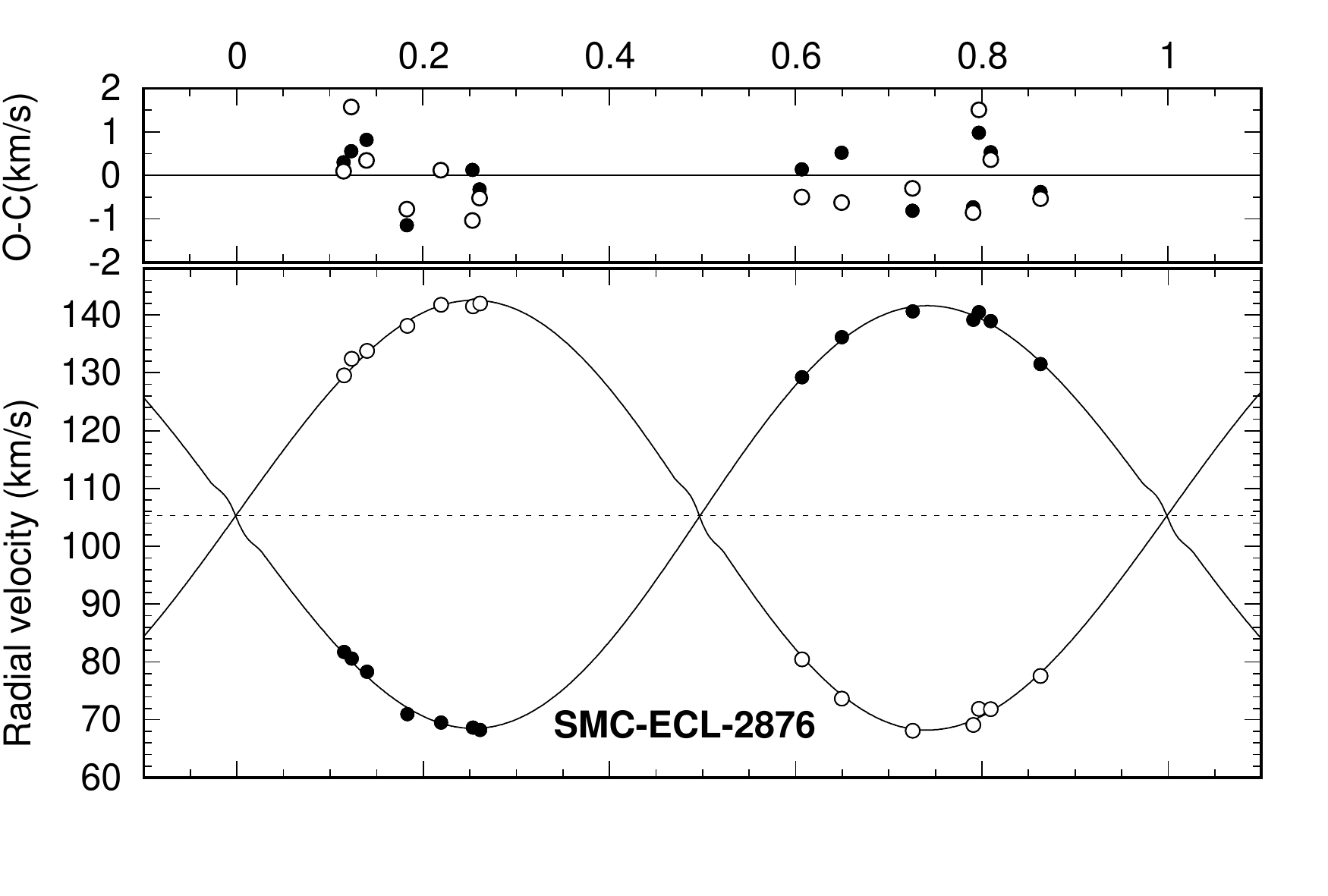} \vspace{-1.0cm}
\mbox{}\\ 
\includegraphics[angle=0,scale=.47]{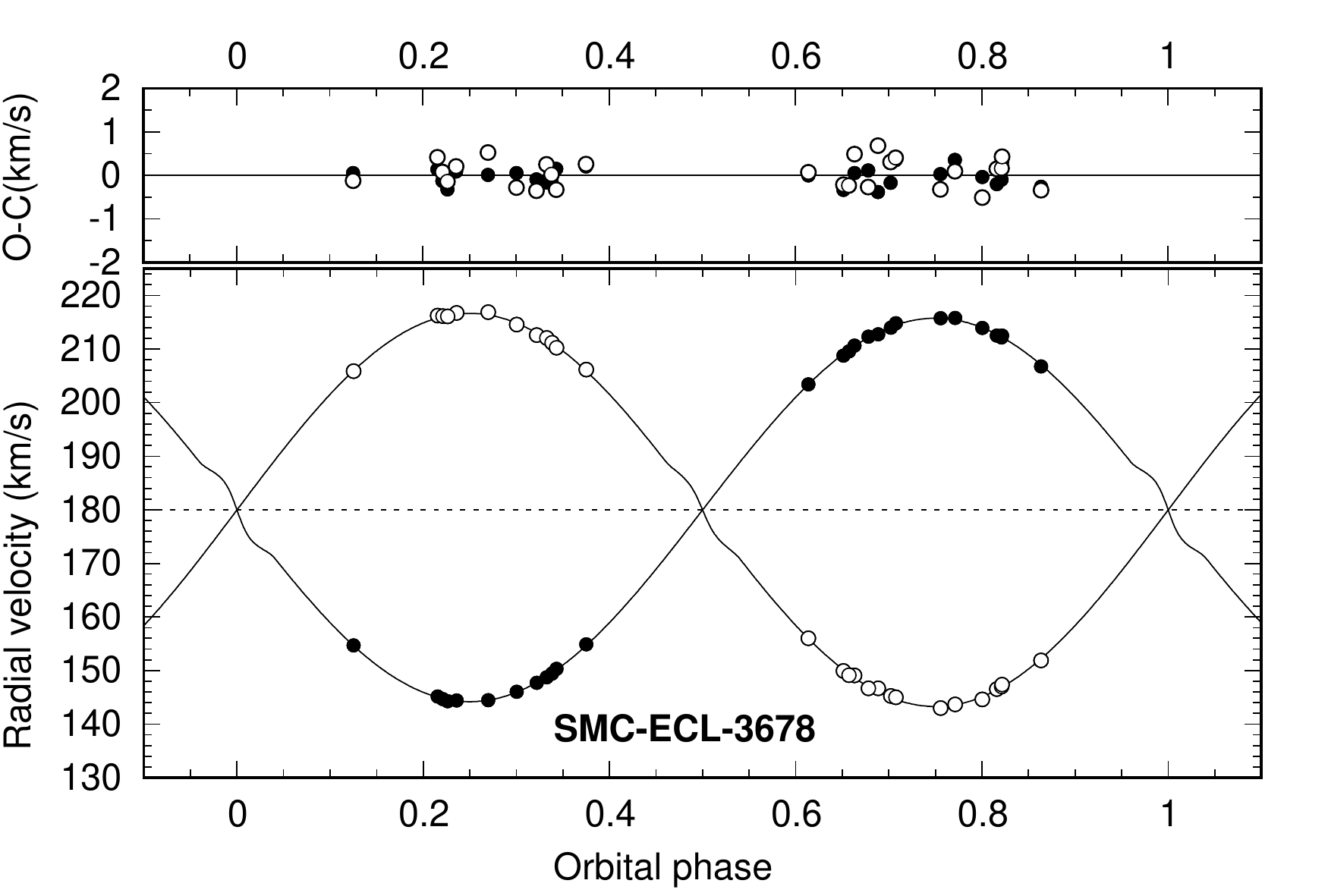}
\end{minipage}\hfill 
\caption{ The radial velocity curve solutions to ten eclipsing binaries in the SMC.\label{fig1}}
\end{figure*}

The third method uses the effective temperatures derived from the atmospheric analysis described in Sec.~\ref{sec:atmo}. We estimated the intrinsic ($V\!-\!I$) and ($V\!-\!K$) colors of each component from  effective temperature - ($V\!-\!I$) and ($V\!-\!K$) color calibrations \citep{ben98, hou00, ram05, gon09, wor11}. These colors were compared with the observed colors of the components obtained from the preliminary solution in order to derive E($V\!-\!I$) and E($V\!-\!K$) color excesses. Both excesses were combined to derive a E($B\!-\!V$) color excess. The reddening to each system was calculated as the mean value of the two components with the exception of SMC-ECL-0019 and SMC-ECL-0727, for which we have a reddening estimate from only one component. In these two cases we assigned a half weight to the estimates. For SMC-ECL-1859, however, this procedure leads to an unrealistically large E($B-V$) and we dropped this estimate. These reddening values are reported in fourth column of Table~\ref{tbl-5} and the fifth column presents the final adopted reddening which is the  average of the individual E($B\!-\!V$) estimates for each system.

\subsection{Effective Temperatures}
\label{sec:temp}
The individual extinction estimates were used to calculate ($V\!-\!I$) and ($V\!-\!K$) intrinsic colors of the components using preliminary light curve solutions. We then utilized a number of temperature-color calibrations (see previous Section) to derive color temperatures of the components. On average the color temperatures are smaller  than spectroscopic temperatures by only about 60 K, but in several cases the differences are significantly larger (up to about 600 K), most notably for the primary component of SMC-ECL-1492. In order to perform  an internal cross-checking of radiative parameters we also calculated the bolometric corrections. These were calculated as the average of several calibrations \citep{cas10,mas06,alo99,flo96} assuming  surface temperatures equal to the color temperatures from Table~\ref{tbl:atmo}. 

\subsection{Fitting  Light and Radial Velocity Curves} 
\label{fit}
 The light and radial velocity curves were solved simultaneously following the methodology described by G14. We used the Wilson-Devinney code version 2007 \citep{wil71,wil79,wil90,van07} equipped with a Python wrapper.  Light curves and radial velocity curves were weighted according to the standard deviation of their residuals. The weights were updated a few times in the course of the analysis. We used a logarithmic law (\verb"LD="$-$\verb"2" for both stars) for the limb darkening. Limb darkening coefficients were calculated automatically and updated in every iteration using grids of coefficients from \cite{van93}. We searched for the solution giving the best $\chi^2$. We set values of the control parameters as: the operation mode \verb"MODE=2" (detached configuration), \verb"IPB=0", the simple reflection treatment \verb"MREF=1", numerical grid size \verb"N1=N2=40", coarse grid size \verb"20", and the atmospheric model approximation \verb"IFAT1=IFAT2=1". We used no spots in modeling of the light curves. The relative parameter increments used by the DC module were set to \verb"DEL=0.007" and the Marquardt multiplier was set to \verb"XLAMDA"$=1\cdot10^{-5}$.
 
Initial parameters were found by a trial-and-error approach. In the analysis we adjusted the observed orbital period $P_{\rm obs}$, the epoch of the primary minimum $T_0$, the mass ratio $q$, the semimajor axis $a$, the systemic velocity $\gamma$, the eccentricity $e$, the longitude of periastron $\omega$, the orbital inclination $i$, the effective temperature $T_{\rm eff}$, the modified surface potentials $\Omega_1$ and $\Omega_2$, the relative luminosity of the primary component in a given band $L_1$, and the third light $l_3$.

We included the spectroscopic light ratios in the $V$- and $R_C$-bands in the modeling. We searched for solutions giving a model light ratio in agreement with the spectroscopic light ratios (Table~\ref{tbl:spec2}) to within the estimated errors. For several systems showing partial eclipses we had to enforce solutions by fixing one of the modified Roche lobe potentials $\Omega$, which corresponds approximately to fixing one of the fractional radii $r$, during the fitting procedure. We usually fixed the potential of the smaller component. For SMC-ECL-1492, which shows notable out-of-eclipse variations (because of a significant tidal deformation of the secondary), we additionally fitted albedo parameters ($A_1$, $A_2$). In this case, however, we noted no improvement in $\chi^2$.
 
\begin{deluxetable*}{@{}rlllllll | lllcc}
\tabletypesize{\scriptsize}
\tablecaption{Model parameters from the Wilson-Devinney code \label{tbl-3}}
\tablewidth{0pt}
\tablehead{
\colhead{OGLE ID} &\multicolumn{7}{c}{Orbital parameters} & \multicolumn{5}{c}{Photometric parameters} \\
\colhead{SMC-ECL-} &\colhead{$P_{\rm orb}$} &\colhead{$q=\frac{M_2}{M_1}$}&\colhead{$a$} &\colhead{$\gamma$}&\colhead{$e$} & \colhead{$\omega$} &\colhead{$K$} &\colhead{$i$} & \colhead{$T_{\rm eff}$} &\colhead{$r$}&\colhead{$\left(\frac{L2}{L1}\right)_V$} & \colhead{$\left(\frac{L2}{L1}\right)_{{\rm M}\!_B}$}  \\
\colhead{} &\colhead{(days)} &\colhead{}&\colhead{($R_\odot$)}& \colhead{(km s$^{-1}$)}&\colhead{} & \colhead{(deg)} &\colhead{(km s$^{-1}$)}& \colhead{(deg)} & \colhead{(K)} & \colhead{} & \colhead{$\left(\frac{L2}{L1}\right)_I$} & \colhead{$\left(\frac{L2}{L1}\right)_{{\rm M}\!_R}$} \vspace{0.08cm}
}
\startdata
 0019$\;$ p & 143.9391(19)& 0.979(11)& 174.2(10)& 122.23(20) & 0 & -- & 30.22(24) &  85.87(21) & 5200 &  0.0964(21) &  1.14 & -- \\
            s & & & & & & &30.86(26)  & & 5105 & 0.1079(34) &  1.17 & -- \\
    0439$\;$ p & 279.2663(42)& 1.009(7)& 298.3(11)& 127.17(9) &  0.356(3) & 202.8(4) & 29.03(13) &  88.36(6) & 5250 &  0.0486(15) &  1.15 & -- \\
            s & & & & & & &28.77(15)  & & 5265 & 0.0516(16) &  1.14 & -- \\
    0727$\;$ p & 316.5722(60)& 0.960(9)& 339.8(16)& 162.70(12) &  0.299(3) & 130.0(5) & 27.88(20) &  89.57(16) & 5300 &  0.0431(5) &  1.17 &  1.14 \\
            s & & & & & & &29.03(18)  & & 4835 & 0.0599(7) &  1.34 &  1.28 \\
    0970$\;$ p & 191.5510(14)& 0.983(6)& 198.5(7)& 143.37(8) &  0.368(3) & 121.3(3) & 27.94(11) &  88.41(16) & 4850 &  0.0927(13) &  1.21 &  1.21 \\
            s & & & & & & &28.42(14)  & & 4780 & 0.1065(12) &  1.24 &  1.23 \\
   1492$\;$ p &  73.7150(4)& 1.001(9)&  92.3(4)& 169.35(11) & 0 & -- & 30.71(19) &  75.60(37) & 4980 &  0.1725(61) &  1.94 &  1.92 \\
            s & & & & & & &30.68(20)  & & 4775 & 0.2714(28) &  2.09 &  2.04 \\
   1859$\;$ p &  75.5436(3)& 0.960(6)& 129.1(4)& 146.55(10) & 0 & -- & 42.24(12) &  86.30(7) & 5550 &  0.1253(24) &  0.73 &  0.71 \\
            s & & & & & & &44.01(25)  & & 4950 & 0.1450(21) &  0.87 &  0.82 \\
   2761$\;$ p & 150.3639(8)& 0.978(9)& 201.0(9)& 153.80(11) &  0.091(4) &  95.6(3) & 33.52(18) &  86.95(5) & 5630 &  0.0860(14) &  0.98 & -- \\
            s & & & & & & &34.29(24)  & & 5580 & 0.0871(19) &  1.00 &  0.99 \\
   2876$\;$ p & 120.8666(12)& 0.984(9)& 176.7(8)& 105.24(16) &  0.020(6) & 259(3) & 36.55(21) &  84.96(8) & 5560 &  0.0935(21) &  0.74 & -- \\
            s & & & & & & &37.16(26)  & & 5150 & 0.0979(20) &  0.83 & -- \\
   3529$\;$ p & 234.3260(12)& 1.052(11)& 270.6(15)& 177.53(14) &  0.450(1) & 181.3(7) & 33.48(30) &  86.54(14) & 5835 &  0.0466(16) &  2.06 & -- \\
            s & & & & & & &31.83(18)  & & 5290 & 0.0781(9) &  2.27 &  2.19 \\
   3678$\;$ p & 187.8429(7)& 0.975(2)& 270.8(3)& 179.97(3) & 0 & -- & 35.79(5) &  83.73(6) & 6650 &  0.1152(18) &  0.47 &  0.44 \\
            s & & & & & & &36.72(7)  & & 4980 & 0.1574(16) &  0.75 &  0.64 \\
   1421$\;$ p & 102.8342(4)& 1.045(6)& 163.7(4)& 187.78(8) & 0 & -- & 41.15(16) &  88.05(33) & 5395 &  0.1084(17) &  1.24 &  1.22 \\
            s & & & & & & &39.36(14)  & & 5020 & 0.1456(13) &  1.38 &  1.33                                                     
\enddata 
\tablecomments{Quoted uncertainties are the standard errors from the Differential Corrections subroutine combined with errors from
Monte Carlo simulations with the JKTEBOP code ver.~34 (see Section~\ref{fin} for details).}
\end{deluxetable*}

\section{Adopted solutions and final parameters}
\label{fin}
\begin{figure*}
\begin{minipage}[th]{0.5\linewidth}
\includegraphics[angle=0,scale=.49]{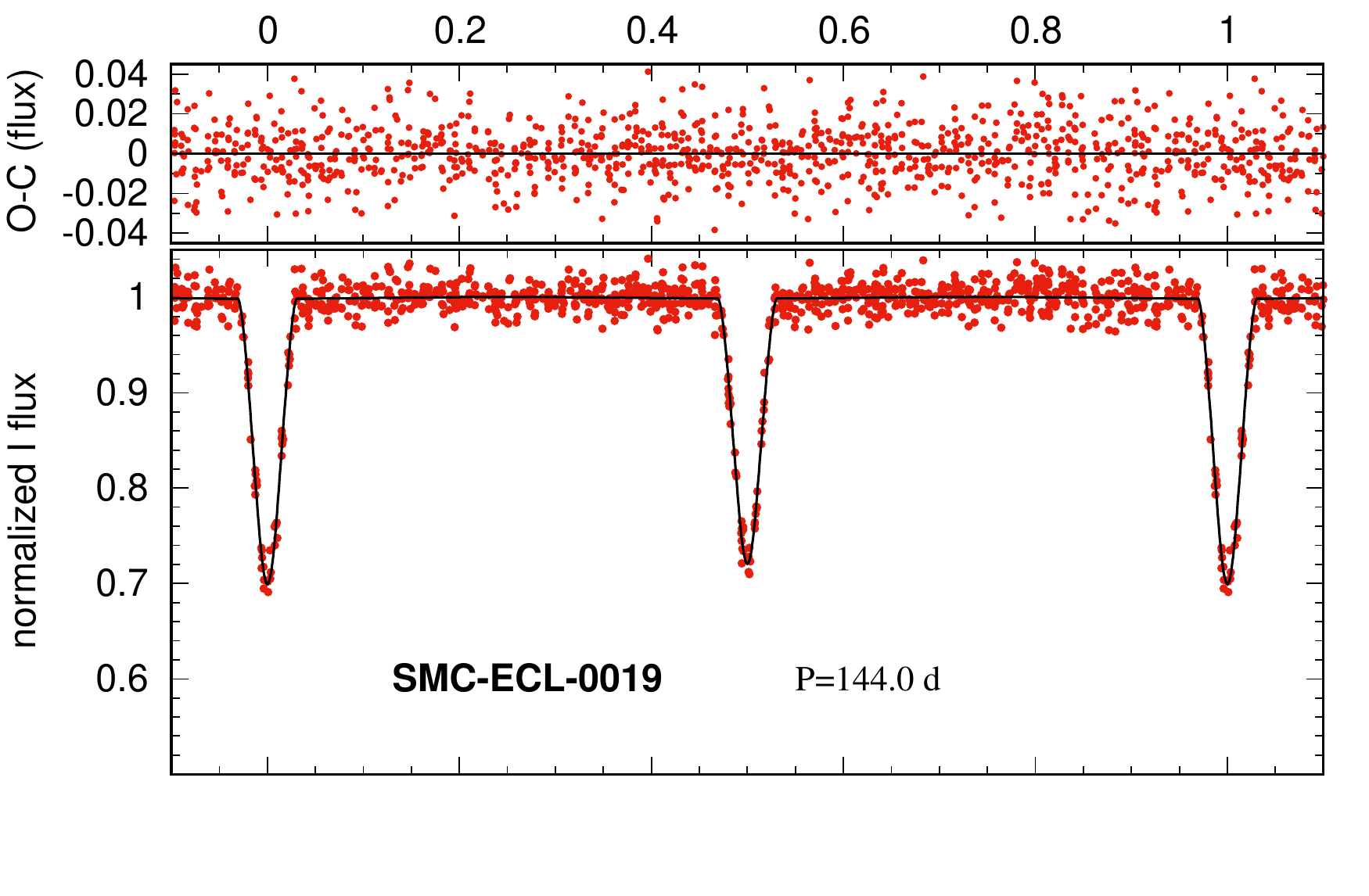} \vspace{-1.17cm}
\mbox{}\\ 
\includegraphics[angle=0,scale=.49]{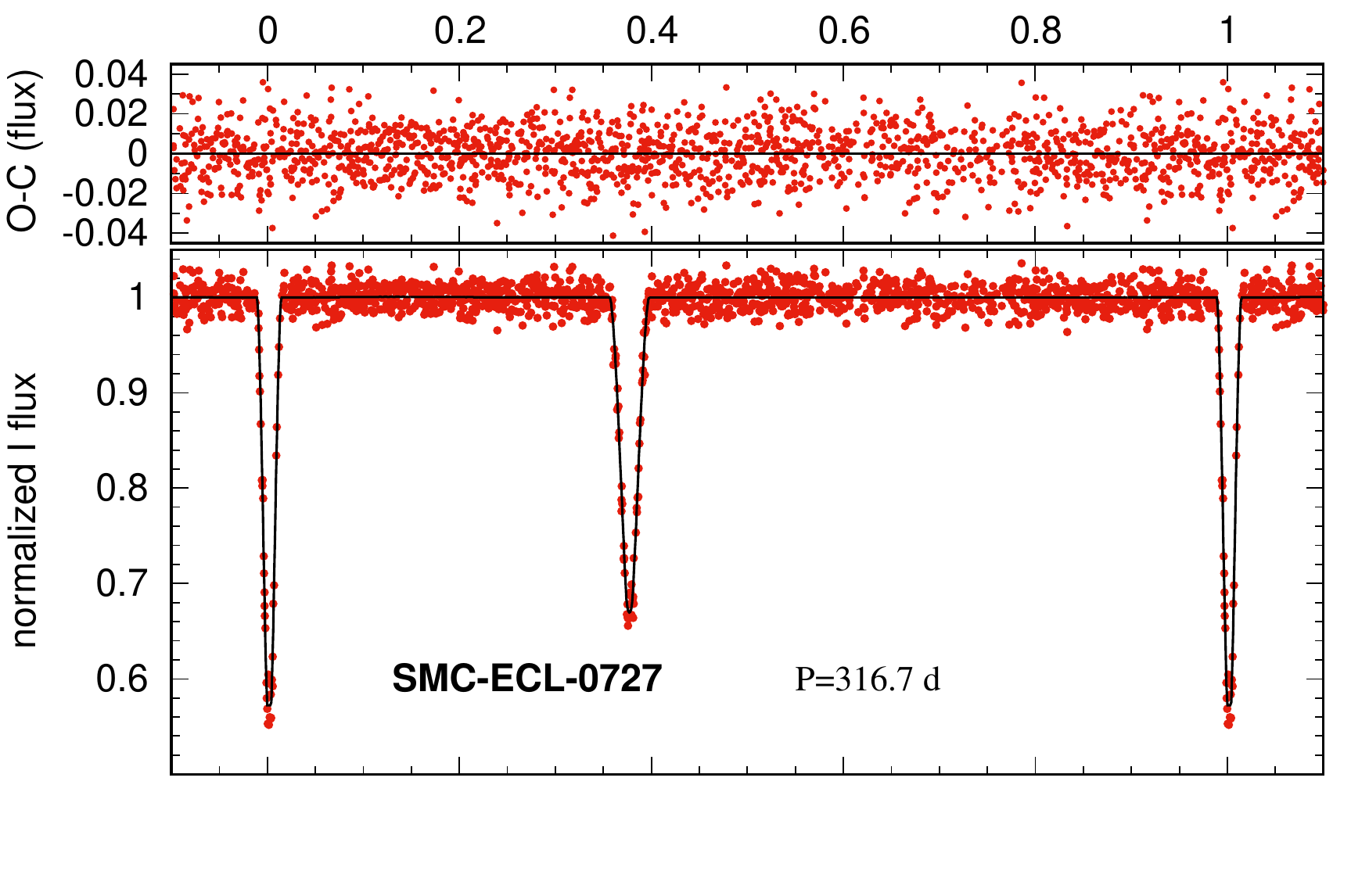}\vspace{-1.17cm}
\mbox{}
\includegraphics[angle=0,scale=.49]{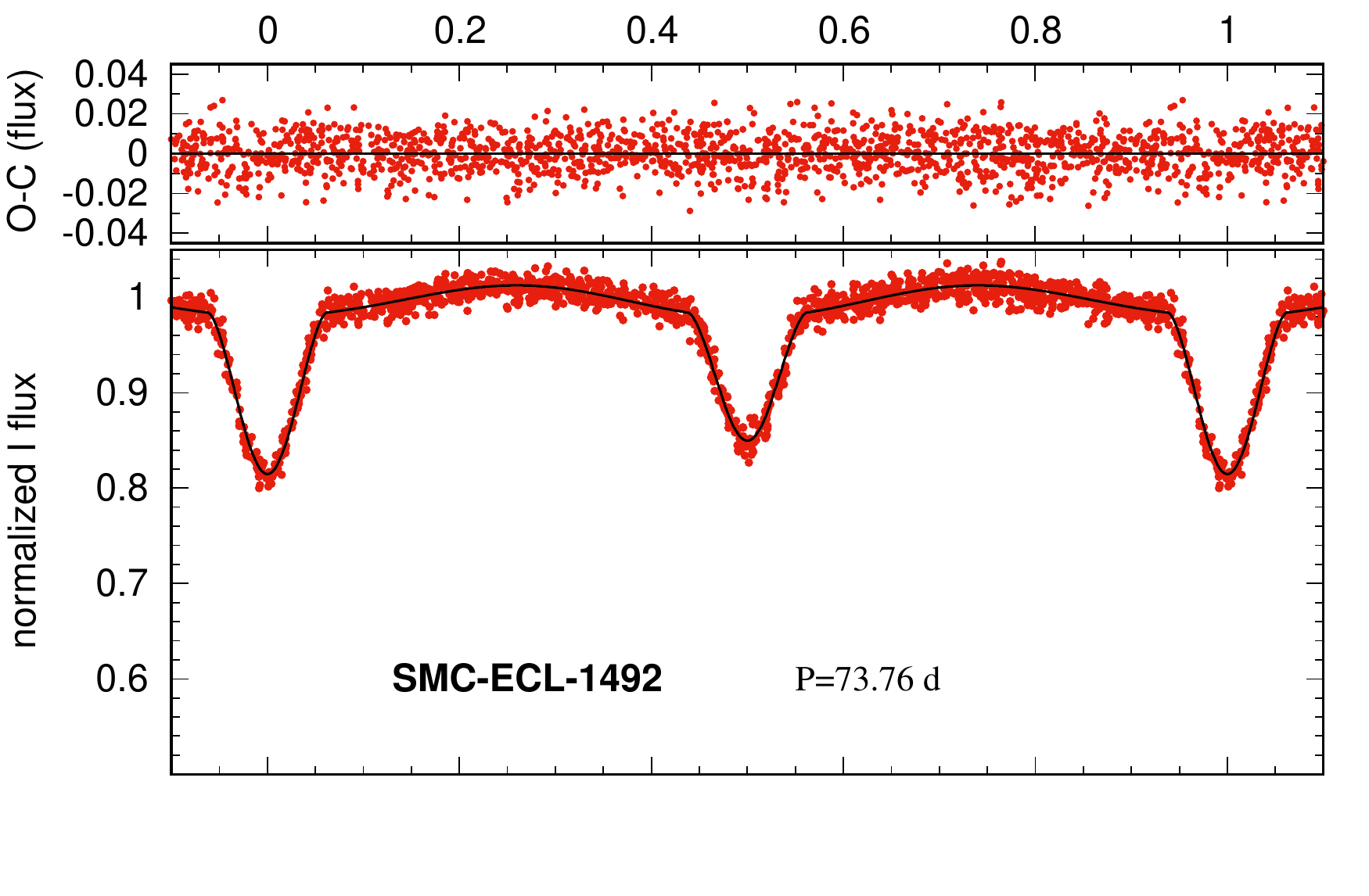}\vspace{-1.17cm}
\mbox{}
\includegraphics[angle=0,scale=.49]{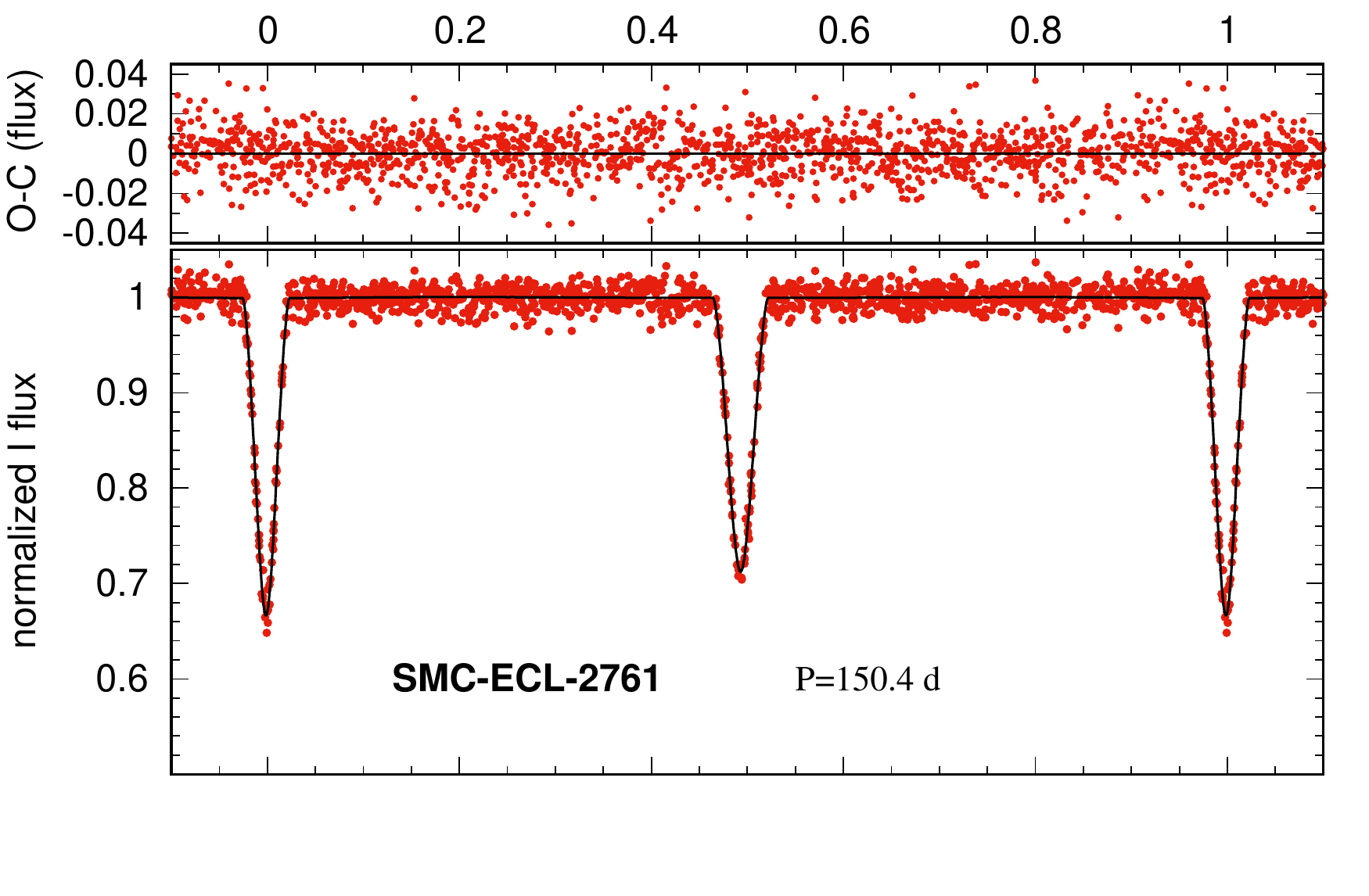} \vspace{-1.17cm}
\mbox{}\\ 
\includegraphics[angle=0,scale=.49]{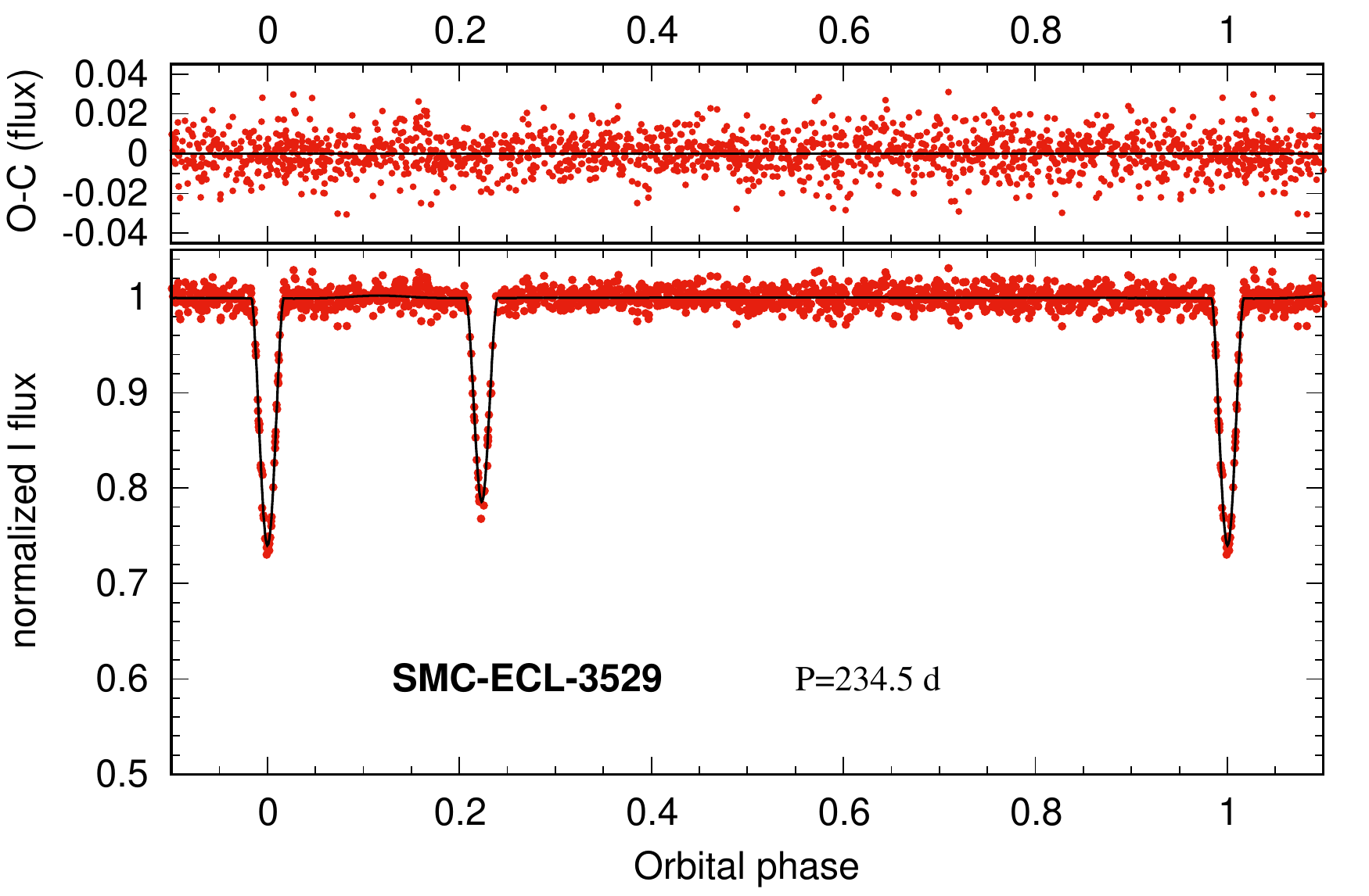}
\end{minipage}\hfill 
\begin{minipage}[th]{0.5\linewidth}
\includegraphics[angle=0,scale=.49]{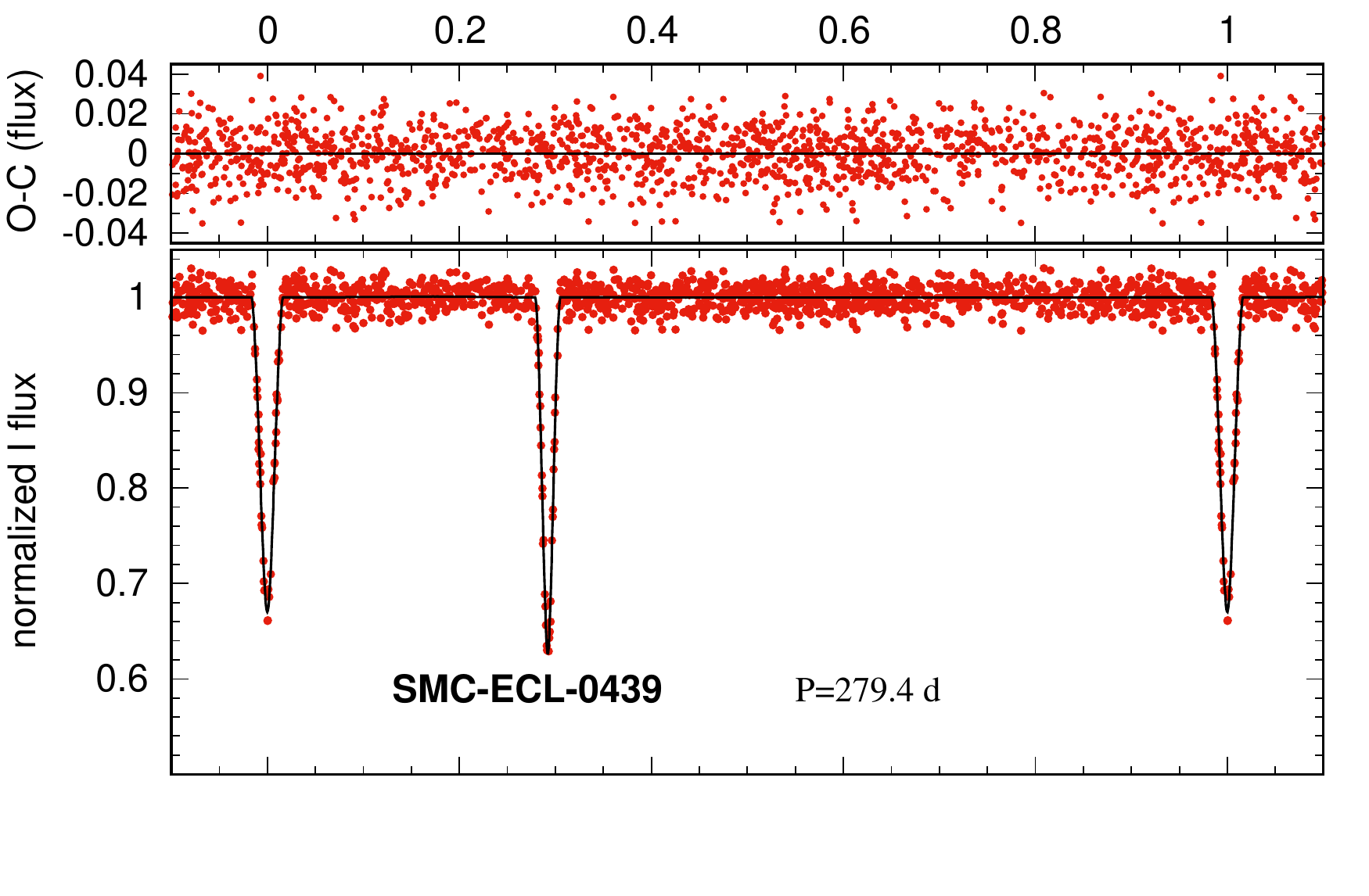}\vspace{-1.17cm}
\mbox{}
\includegraphics[angle=0,scale=.49]{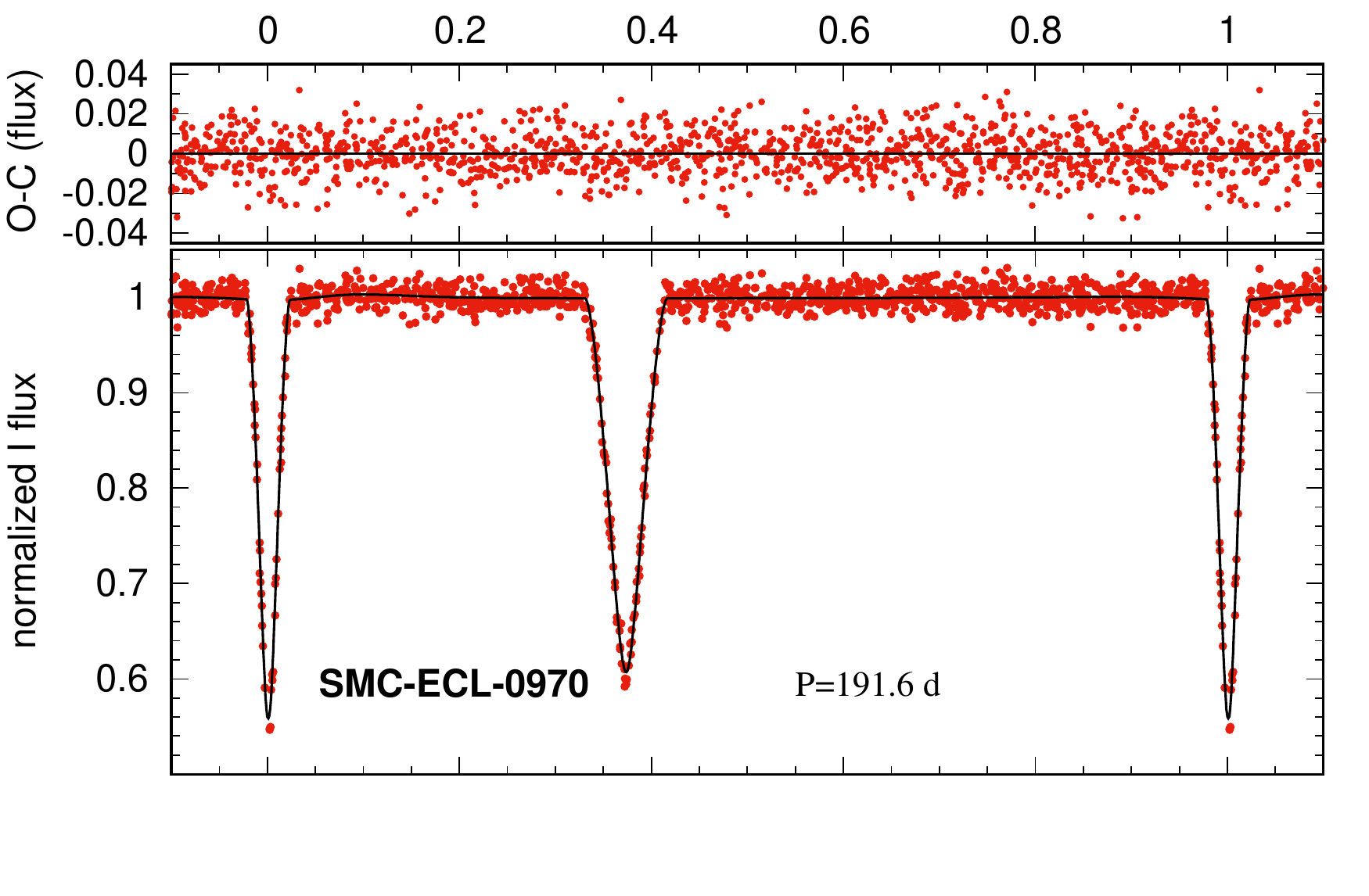}\vspace{-1.17cm}
\mbox{}
\includegraphics[angle=0,scale=.49]{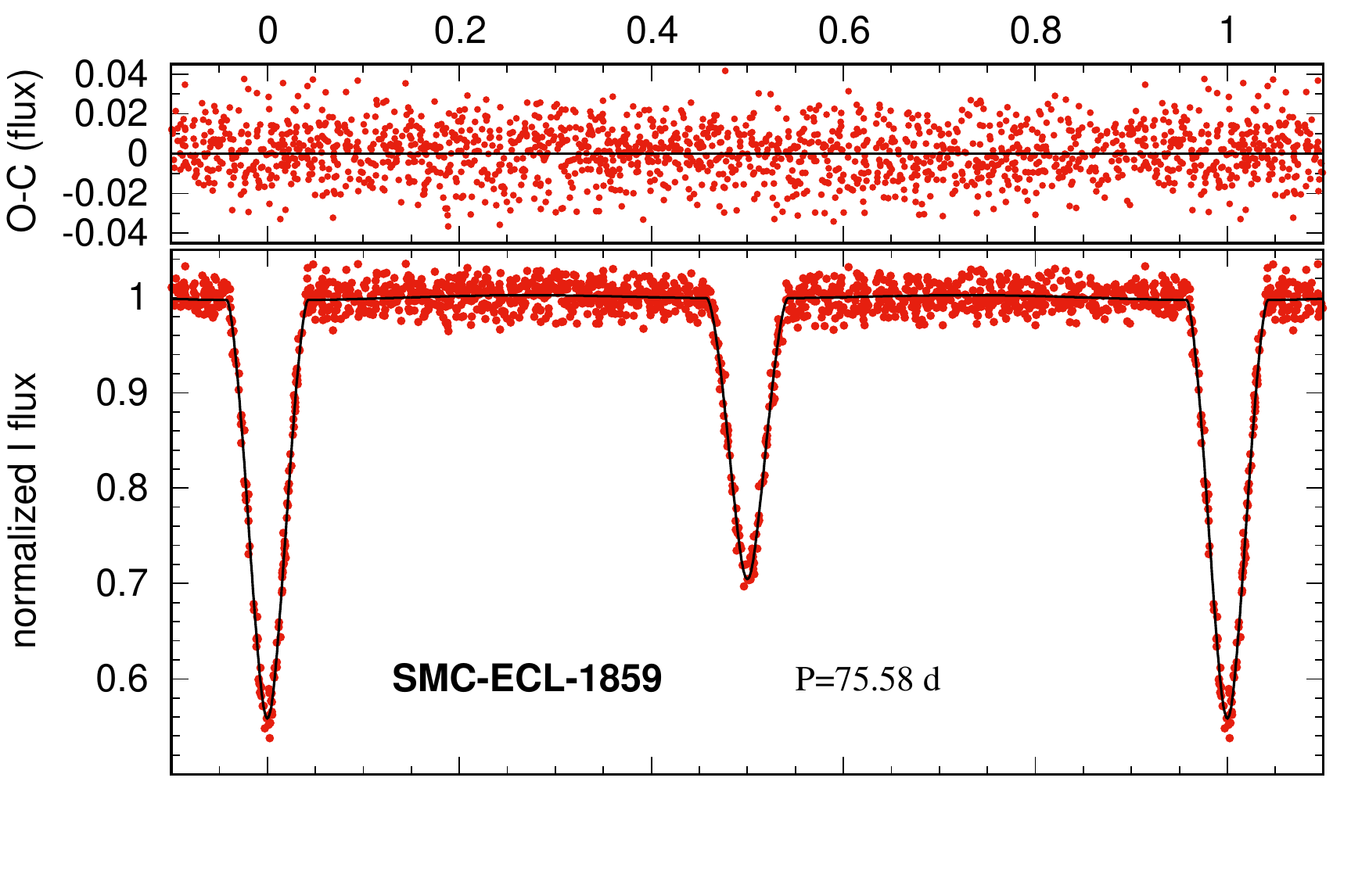}\vspace{-1.17cm}
\mbox{}
\includegraphics[angle=0,scale=.49]{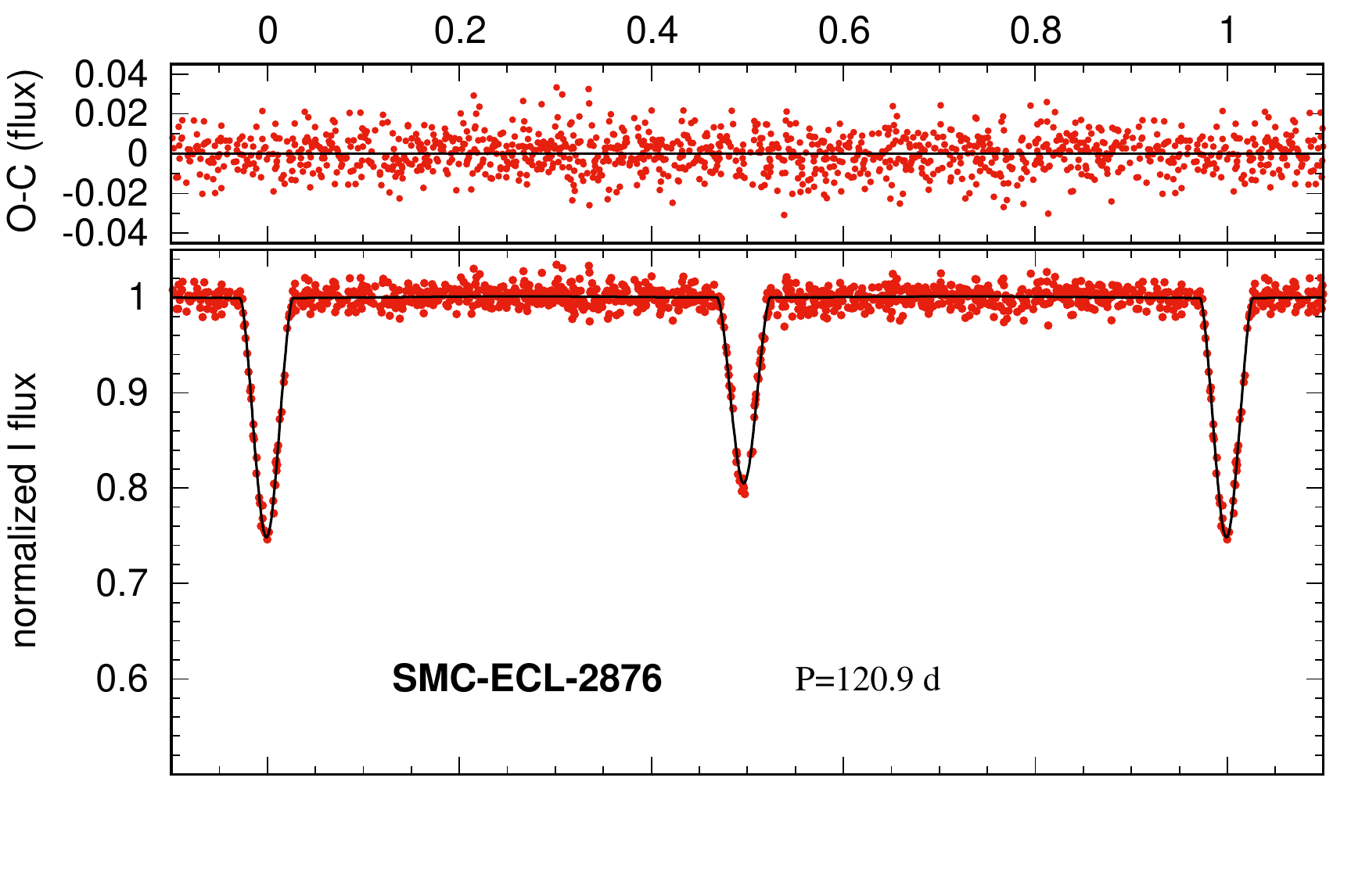} \vspace{-1.17cm}
\mbox{}\\ 
\includegraphics[angle=0,scale=.49]{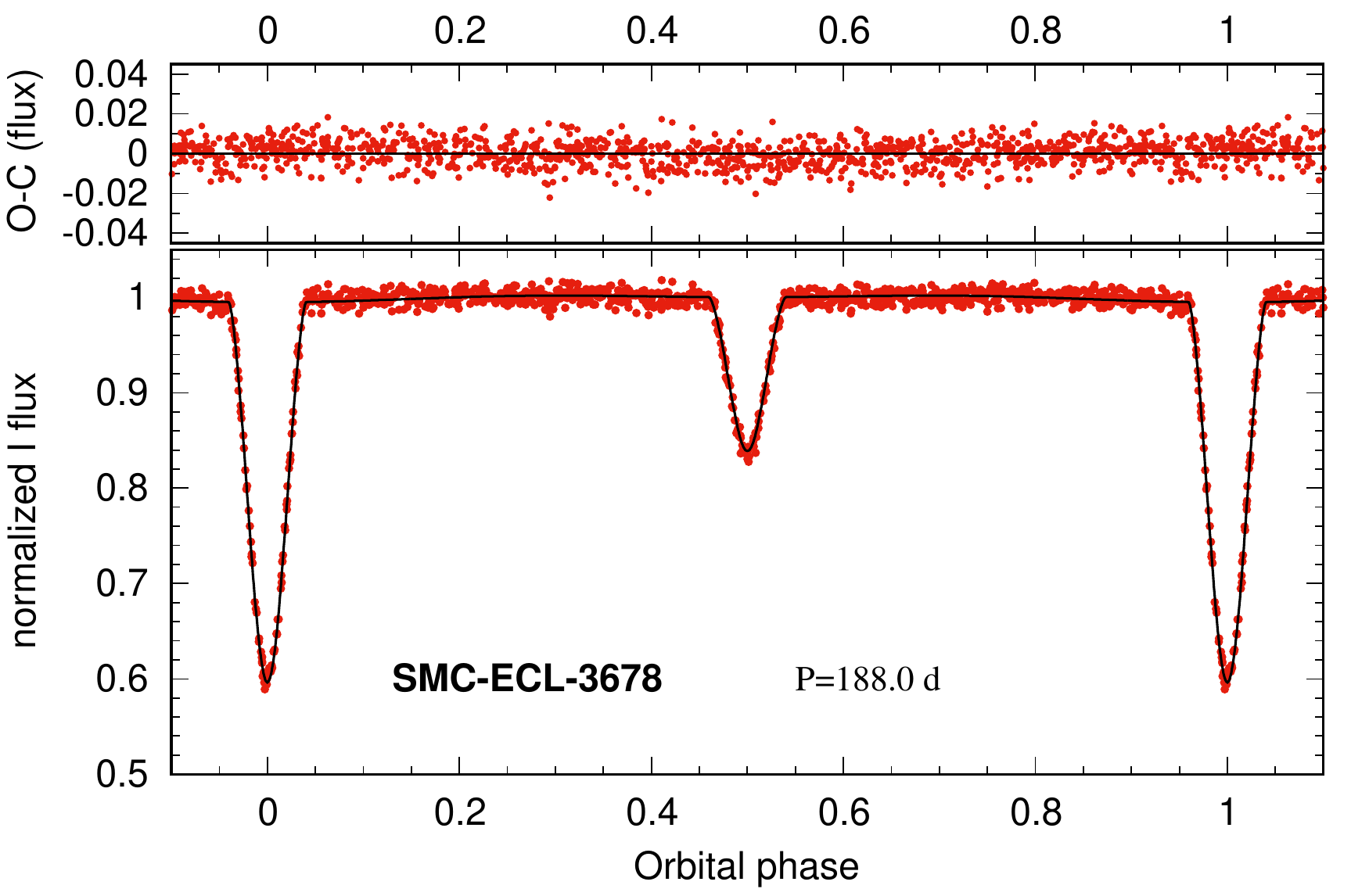}
\end{minipage}\hfill 
\caption{ The I$_{\rm C}$-band light curve solutions to ten eclipsing binaries in the SMC.\label{fig2}}
\end{figure*}

We computed two types of models for each system: 1) setting limb darkening coefficients according to a logarithmic law (\verb"LB"$=-2$) and the third light $l_3=0$; and 2) adjusting the third light $l_3$ in all bands with a logarithmic law of limb darkening.  We compared the models according to their reduced $\chi^2_{\rm r}$ and chose the one with the lower value. We confirmed the presence of  third light in only one case, SMC-ECL-1421  (see also G14). For the case of SMC-ECL-1492 we also suspect  a blue third light. Table~\ref{tbl-3} lists the parameters of the best model for each system. The Table shows true orbital periods $P_{\rm orb}=P_{\rm obs}/(1+\gamma/c)$, where $P_{\rm obs}$ are given in Table~\ref{tbl:1} and $c$ is the speed of light, $K$ is the radial velocity semiaplitude, $r$ is the fractional radius and $(L_2/L_1)$ are light ratios in the $V$, $I_C$, MACHO $B$ and MACHO $R$ bands. The meaning of the other symbols is explained in Section~\ref{fit}. Individual model solutions to the radial velocity curves are presented in Fig.~\ref{fig1} and the $I$-band light curve solutions are presented in Fig.~\ref{fig2}. 

The quoted uncertainties were calculated by combining the formal errors reported by the Differential Correction routine of the Wilson-Devinney code and errors from the Monte-Carlo simulations with the JKTEBOP code ver. 34 \citep{pop81,sou04,sou13}. We adopted the larger error of the two. For the JKTEBOP Monte Carlo simulations   we ran 10000 models, simultaneously solving light and radial velocity curves. For the JKTEBOP calculations we used the same set of free parameters as for the WD code. Very good agreement (to within of $2\sigma$) was usually obtained between model parameters of the WD code and the JKTEBOP code, not surprising because all systems but one are well detached. Only in the case of SMC-ECL-1492  was the disagreement on the level of $3.5\sigma$. This is because the secondary is significantly distorted and the simple geometry provided by JKTEBOP is not fully sufficient.

Absolute physical parameters for all components are given in Table~\ref{tbl-6}. The masses and radii of the stars were calculated by adopting astrophysical constants from IAU resolution B3 \citep{mam15}. Spectral types for the components were indirectly derived from surface temperatures and gravities using the calibration of \cite{alo99}. For all components the accuracy of the mass determinations is better than 2\% and for the radii determinations better than 3.5\%. The metallicity of each system was calculated as a flux weighted average of the metallicity of both components. In this way we assumed the same composition for the components in any one system.    

\subsection{Notes on Individual Systems}
In this paper we focus on a determination of the distance to the Small Magellanic Cloud, and a more extensive discussion of individual systems, including their evolutionary status in a manner similar to that done for the LMC eclipsing binaries \citep{gra18}, will be presented in a subsequent paper. Here we present a brief description of three systems due to their unusual features.

\subsubsection{OGLE SMC-ECL-1492}
This system consists of two giant stars of equal masses but significantly different radii. The masses of the components are sub-solar, indicative of a large age of the components (age $\gtrsim$ 10 Gyr). However, the small separation of the components may have allowed  mass transfer between the components during the evolutionary phases of ascending the red giant branch and significant mass loss from the system. In this case the age of the system would be  smaller. Our analysis shows that a blue third light may be present in the system. The detection is quite weak (1.5$\sigma$) and inclusion of the third light does not significantly improve the solution. If the third light is real its brightness and colors would roughly correspond to an early A-type main sequence star. Such a star should have a mass of about 2 $M_\odot$, and  thus  could not coevolve with the components of the binary. For the purpose of this work we assumed no companion or optical blend in the system, but we included additional uncertainty on the third light in the error budget of its distance.

\begin{deluxetable*}{@{\hspace{-12pt}}rccclllllLLLl}
\tabletypesize{\scriptsize}
\tablecaption{Physical Properties of Eclipsing Binary Giants in the SMC \label{tbl-6}}
\tablewidth{0pt}
\tablehead{
\colhead{OGLE ID} &\colhead{Spectral}&\colhead{V$_0$}& \colhead{(V$-$K)$_0$}&\colhead{Mass}& \colhead{Radius}&\colhead{$\log{g}$}& \colhead{$T_{\rm eff}$}&\colhead{Lumin.} &\colhead{$M_{\rm bol}$}&\colhead{$M_V$}&\colhead{$\left[{\rm Fe/H}\right]$}&\colhead{$v\sin{i}$}\\
\colhead{SMC-ECL-}&\colhead{Type} & \colhead{(mag)} &\colhead{(mag)} & \colhead{M$_\sun$} &\colhead{R$_\sun$} & \colhead{(cgs)} &\colhead{(K)} &\colhead{L$_\sun$}&\colhead{(mag)}&\colhead{(mag)}&\colhead{(dex)}&\colhead{km s$^{-1}$}
}
\startdata
   0019$\;$ p & G4III & 18.52(9)&  2.01(9)& 1.73(3) & 16.8(4) &  2.23(2) & 5200(100) &   186(17) & -0.92(10) & -0.63(9) & -0.56(8) &   7.9(0.7) \\
            s & G5III & 18.38(9) & 2.09(9) & 1.69(3) & 18.8(6) &  2.12(3) & 5105(100) &   216(22) & -1.09(11) &-0.77(9) & &   8.7(0.8)\\
   0439$\;$ p & G2III & 18.72(5)&  1.95(5)& 2.27(3) & 14.5(5) &  2.47(3) & 5250(100) &   144(14) & -0.64(11) & -0.39(5) & -1.25(8) &   3.1(0.9) \\
            s & G2III & 18.58(5) & 1.93(5) & 2.29(3) & 15.4(5) &  2.42(3) & 5265(100) &   164(16) & -0.79(11) &-0.54(5) & &   3.5(1.6)\\
   0727$\;$ p & G3III & 18.59(9)&  1.86(9)& 2.68(4) & 14.6(2) &  2.53(1) & 5300(100) &   152(12) & -0.71(9) & -0.52(9) & -0.89(8) &   2.6(1.6) \\
            s & G8III & 18.42(9) & 2.28(9) & 2.57(4) & 20.4(3) &  2.23(1) & 4835(90) &   204(16) & -1.02(9) &-0.69(9) & &   5.0(1.6)\\
   0970$\;$ p & G8III & 18.54(5)&  2.29(5)& 1.44(2) & 18.4(3) &  2.07(1) & 4850(70) &   169(11) & -0.82(7) & -0.45(5) & -0.83(8) &   6.4(1.2) \\
            s & G9III & 18.33(5) & 2.36(5) & 1.42(1) & 21.1(2) &  1.94(1) & 4780(70) &   210(13) & -1.06(7) &-0.65(5) & &  12.0(1.2)\\
   1492$\;$ p & G6III & 18.67(4)&  2.16(4)& 0.97(1) & 15.9(6) &  2.02(3) & 4980(100) &   141(15) & -0.62(12) & -0.31(4) & -0.88(9) &   9.3(1.2) \\
            s & G8III & 17.95(4) & 2.36(4) & 0.97(1) & 25.1(3) &  1.63(1) & 4775(70) &   294(19) & -1.42(7) &-1.03(4) & &  11.2(0.5)\\
   1859$\;$ p & G0III & 18.10(4)&  1.69(4)& 2.58(3) & 16.2(3) &  2.43(2) & 5550(100) &   224(18) & -1.12(9) & -0.96(4) & -0.21(22) &   9.5(1.2) \\
            s & G7III & 18.44(4) & 2.19(4) & 2.48(2) & 18.7(3) &  2.29(1) & 4950(100) &   189(16) & -0.94(9) &-0.62(4) & &  13.0(1.1)\\
   2761$\;$ p & F9III & 17.79(4)&  1.64(5)& 2.44(4) & 17.3(3) &  2.35(1) & 5630(120) &   270(25) & -1.33(10) & -1.18(4) & -0.73(9) &   8.9(1.0) \\
            s & G0III & 17.80(4) & 1.67(5) & 2.38(3) & 17.5(4) &  2.33(2) & 5580(120) &   268(26) & -1.32(11) &-1.16(4) & &   8.3(0.8)\\
   2876$\;$ p & G0III & 17.92(4)&  1.65(4)& 2.56(4) & 16.5(4) &  2.41(2) & 5560(100) &   235(20) & -1.18(9) & -1.06(4) & -0.43(8) &   7.2(0.7) \\
            s & G4III & 18.25(4) & 1.98(4) & 2.51(3) & 17.3(4) &  2.36(2) & 5150(90) &   190(15) & -0.95(9) &-0.73(4) & &   8.1(0.7)\\
   3529$\;$ p & F8III & 18.21(9)&  1.49(9)& 2.36(3) & 12.6(4) &  2.61(3) & 5835(100) &   166(16) & -0.80(11) & -0.69(9) & -0.93(10) &   9.1(0.8) \\
            s & G3III & 17.42(9) & 1.74(9) & 2.48(5) & 21.1(3) &  2.18(1) & 5290(180) &   315(44) & -1.50(15) &-1.48(9) & &   7.7(1.1)\\
   3678$\;$ p &  F2II & 15.77(6)&  1.05(6)& 3.83(2) & 31.2(5) &  2.03(1) & 6650(130) &  1715(145) & -3.34(9) & -3.21(6) & -0.78(8) &   7.9(2.1) \\
            s &  G6II-III & 16.59(6) & 2.21(6) & 3.73(1) & 42.6(4) &  1.75(1) & 4980(70) &  1007(60) & -2.76(7) &-2.39(6) & &  11.3(0.9)\\
   1421$\;$ p & G2III& 17.93(6)&  1.79(6)& 2.72(2) & 17.7(3) &  2.37(1) & 5395(100) &   240(19) & -1.20(9) & -1.03(6) & -1.05(11) &  11.2(1.1) \\
            s &G6III & 17.70(6) & 2.10(6) & 2.85(2) & 23.8(2) &  2.14(1) & 5020(95) &   325(25) & -1.53(8) &-1.27(6) & &  11.8(0.5)
 \enddata
 \tablecomments{Absolute dimensions were calculated assuming nominal solar gravitational constant $\mathcal{GM}_\sun=1.3271244\!\cdot\!10^{20}$ m$^3$ s$^{-2}$, nominal solar radius $\mathcal{R}_\sun=695 700$ km, solar effective temperature $\mathcal{T}_{\rm eff,\sun}=5772$ K \citep{mam15}, and the solar bolometric absolute magnitude $M_{bol,\sun}=+4.75$. The magnitudes and colors are extinction corrected values. The $(V\!-\!K)_0$ color is expressed in the 2MASS system.}
\end{deluxetable*}

\subsubsection{OGLE SMC-ECL-3678}
This is the brightest and most massive binary in the sample. The system consists of two unequal components with a large difference in surface temperature. The eclipses are partial but the light curves are of high quality, allowing for an accurate determination of the absolute dimensions. The third light is consistent with $l_3=0$ in all bands and both stars rotate synchronously. The physical parameters of the hotter primary star place it close to the blue edge of the instability strip of the classical Cepheids. To within the precision of the OGLE photometry (0.01 mag) we could not detect any pulsations in the $I$-band light curve. We used the pulsation code from \citet{smo08} to investigate the position of this star with respect to the instability strip of the first overtone Cepheids with metallicity $-0.78$ dex. The star is 2.3$\sigma$ too hot or 6$\sigma$ too bright to fall within the strip. In addition to within a 1$\sigma$ uncertainty in metallicity all pulsation modes are not excited, in agreement with observations.  

\subsubsection{OGLE SMC-ECL-1421 (SMC 101.8 14077)}
The system has been analysed by our team (G14), however we have reinvestigated the possible presence of  third light by including the  MACHO $B$- and $R$-band photometry -- source MACHO 208.15855.1085 \citep{fac07}. The simultaneous solution to the light curves in four bands leads to an unequivocal detection of blue third light in the system, contributing approximately 8\% and 4\% of the light in the $V$- and $I_C$-band, respectively. 
The third light ($V_0$=19.80 mag, $(V\!-\!I_C)_0=0.07$ mag) is fully consistent with a normal A0.5V main sequence star in the SMC \citep{pec13} assuming a system reddening of $E(B\!-\!V)=0.067$ mag. Because such a star should have a mass of about 2.2 M$_\odot$ (smaller than the individual masses of the components of SMC-ECL-1421) we can interpret the third light as indicating an unevolved physical blue companion to the system. The contribution of the third light to the $K$-band is about 1.4\%. The presence of the blue tertiary makes the intrinsic colors of the components redder and thus their temperatures lower by about 150 K, in better agreement with the spectroscopically determined temperature of the secondary (G14). Other physical parameters of the components, for example the  masses and radii, are only marginally different from our previous determinations. The resulting errors on the parameters are reduced.

\section{Distance determination to the SMC}
\subsection{Calculation of Individual Eclipsing Binary Distances}
\label{dist:ind}
The distance to each eclipsing binary was calculated using the surface brightness-color (SBC) calibration for late-type giant stars done by interferometry in the near-infrared domain and presented by \cite{pie19}. This SBC calibration was already used to determine the LMC distance and permits a measurement of the angular diameters of giant stars with an accuracy of up to 0.8\% \citep[see also][]{gal18}. The SBC relation by \cite{pie19} was confirmed by interferometric observations in the visible domain \citep{nar20} and it remains consistent with the revised SBC relation by \cite{sal20} based on a careful selection of F5/K7-II/III stars. The ratio of the angular diameter to the physical radius of an eclipsing binary component gives a direct measurement of the distance to the system. Details of the method are given in \cite{pie13} and \cite{gra14}. Using this new SBC relation we also recalculated  distances to five previously published eclipsing binary stars \citep[][G14]{gra12}, leading to a final  sample of 15 stars. The intrinsic colors of the stars were calculated assuming the "average" Galactic extinction curve from \cite{fit07}: $R=3.1$, $A_K=0.26 E(B\!-\!V)$. We adopted the mean of the distances to the individual components as the measured distance to the binary system. For the case of SMC-ECL-3678  the primary is too blue to apply the SBC relation, and the distance to this system is calculated only from the secondary component. 

We adopted a uniform systematic uncertainty of 1.3\% (0.028 mag in distance modulus) for all systems -- see Sec.~\ref{sec:sys}. The statistical uncertainties of the measured  distances have three main sources: uncertainty in the absolute size of the components and their flux ratios $\sigma(R_1,R_2,L_{21})$, photometric errors in the intrinsic $V-$ and $K$-band magnitudes $\sigma(V_0,K_0)$, and the uncertainty in the third light $\sigma(l_3)$.  Individual distance moduli and distances and their uncertainties are summarized in Table~\ref{tab:dist}. The weighted mean of all 15 distances gives a distance modulus to the SMC  of $m-M=18.978\pm0.089$ mag. The large standard deviation is caused by the physical extension of the SMC itself along the line of sight. The positions of the stars in our sample on the sky and their distance moduli are shown in the left panel of Fig.~\ref{fig:DXY}.

\begin{deluxetable*}{lcccccCcl}
\tabletypesize{\scriptsize}
\label{tab:dist}
\tablecaption{The distances of eclipsing binaries in the SMC\label{tbl-7}}
\tablewidth{0pt}
\tablehead{
\colhead{ID} & \colhead{$(m\!-\!M)$} &\colhead{$\sigma({\rm R_1,\,R_2,\,L_{21}})$} &\colhead{$\sigma (V_0,\,K_0)$}& \colhead{$\sigma (l_{3})$}&\colhead{Combined error} & \colhead{Distance} & \colhead{Ref.} &\colhead{Comment}\\
\colhead{OGLE} &\colhead{(mag)} & \colhead{(mag)}&\colhead{(mag)} &\colhead{(mag)} &\colhead{(mag)}& \colhead{(kpc)} &\colhead{} & \colhead{}
}
\startdata
SMC-ECL-0019 & 19.143 & 0.025 & 0.017 &  0.000   & 0.030 & 67.39\pm0.94 & this work & \\
SMC-ECL-0439 & 19.115 & 0.020 & 0.017 &  0.000   & 0.026 & 66.51\pm0.80 &this work & \\
SMC-ECL-0727 & 19.105 & 0.018 & 0.015 &  0.000   & 0.023 & 66.23\pm0.71 & this work &\\
SMC-ECL-0970 & 18.985 & 0.013 & 0.015 &  0.000   & 0.020 & 62.66\pm0.58 & this work &\\
SMC-ECL-1492 & 18.980 & 0.019 & 0.016 &  0.020   & 0.032 & 62.50\pm0.91 & this work &\\
SMC-ECL-1859 & 19.060 & 0.014 & 0.012 &  0.000   & 0.018 & 64.88\pm0.54 & this work &\\
SMC-ECL-2761 & 18.964 & 0.017 & 0.020 &  0.000   & 0.026 & 62.07\pm0.74 & this work &\\
SMC-ECL-2876 & 18.980 & 0.023 & 0.008 &  0.000   & 0.024 & 62.53\pm0.70 & this work &\\
SMC-ECL-3529 & 18.897 & 0.025 & 0.014 &  0.000   & 0.029 & 60.18\pm0.79 & this work &\\
SMC-ECL-3678 & 18.973 & 0.015 & 0.015 &  0.000   & 0.021 & 62.30\pm0.61 & this work &\\
\hline
\multicolumn{8}{c}{Recalculated distances of previously published systems}\\
SMC-ECL-1421 & 18.963 & 0.012 & 0.010 &  0.016   & 0.023 & 62.03\pm0.64 &  2&SMC101.8 14077 \\
SMC-ECL-0195 & 18.895 & 0.019 & 0.015 &  0.000   & 0.024 & 60.13\pm0.68 & 2&SMC130.5 4296 \\
SMC-ECL-0708 & 18.927 & 0.023 & 0.009 &  0.000   & 0.025 & 61.01\pm0.69 & 2 &SMC126.1 210 \\
SMC-ECL-4152 & 19.028 & 0.015 & 0.015 &  0.000   & 0.022 & 63.93\pm0.63 & 2&SMC108.1 14904\\
SMC-ECL-5123 & 18.792 & 0.016 & 0.012 &  0.000   & 0.020 & 57.33\pm0.53 & 1&SMC113.3 4007
\enddata
\tablecomments{Distance moduli determinations to individual targets together with an error budget. The errors quoted are statistical uncertainties and each determination has an additional systematic uncertainty of 0.028 mag \citep{pie19}. {\it References}: 1 -- \citep{gra12}; 2 -- G14. }
\end{deluxetable*}

\subsection{Sky Position of the SMC Center}
The size of our sample is too small to independently  derive  the  position of the SMC center on the sky, nor the global geometry of the SMC. However, by assuming the coordinates of the SMC center we can  obtain a very accurate determination of its distance. The position of the SMC center is still a matter of  debate and somewhat conflicting results have been reported. A primary reason of these discordant results is the disturbed structure of the SMC itself. The kinematic center of the H~I  disc \citep[e.g.][$\alpha_{2000}=15.2\pm0.4$ deg;  $\delta_{2000}=-72.3\pm0.3$ deg]{DiT19} is often regarded as the center of mass of the SMC. However the H~I disc is disturbed and significantly detached from the optical components of the SMC, and stars in general do not follow its rotation field \citep{Mur19,DeL20}. The old population traced by RR Lyr stars has an optical center lying  south-west of the gas disc center \citep[e.g.][$\alpha_{2000}=14.96$ deg;  $\delta_{2000}=-72.86$ deg]{Mur18}, while the young population traced by Classical Cepheids has a center lying even more to the south-west \citep[e.g.][$\alpha_{2000}=12.54$ deg;  $\delta_{2000}=-73.11$ deg]{Rip17}. Interestingly, star counts based on late type stars \citep{Gon09} and  on intermediate age stars \citep{Rub15} both have centroids very close to the center reported by \cite{Rip17}. 

Our sample consists mostly of young and intermediate age stars with a distribution expected to be similar to that of the Classical Cepheids. Indeed, the Ripepi et al. center divides our sample evenly in the N-S direction (8 stars to the North and 7 stars to the South) and in the E-W direction (7 stars to the East and 8 stars to the West), and we adopt this position as the position of the SMC center. Although the formal uncertainties of the centroid coordinates reported by \cite{Rip17} are only 0.01 deg we have conservatively adopted larger uncertainties for the coordinates of the SMC center: $\alpha_0=12.54\pm0.30$ deg, $\delta_0=-73.11\pm0.15$ deg.

\subsection{Coordinate Transformations}
We converted the Cartesian right-handed equatorial coordinate system into a Cartesian left-handed coordinate system with the $Z$ axis pointed in the direction of the SMC center, which lies at a requested distance from the Sun, the $X$ axis antiparallel to the right ascension,  and the $Y$ axis parallel to the declination. 
The components of a position vector $R$ of a star with equatorial coordinates ($\alpha$, $\delta$) laying at a distance $D$ are:
\begin{equation}
R= D\cdot
\left(\begin{array}{c}
\cos{\alpha}\cos{\delta}\\
\sin{\alpha}\cos{\delta}\\
\sin{\delta}
\end{array}\right)
\end{equation}

The transformation matrix $T$ is:

\begin{equation}
\label{ }
T=
\left(\begin{array}{ccc}
\sin{\alpha_0} & -\cos{\alpha_0} & 0 \\
-\sin{\delta_0}\cos{\alpha_0} & -\sin{\delta_0}\sin{\alpha_0} & \cos{\delta_0} \\
\cos{\delta_0}\cos{\alpha_0} & \cos{\delta_0}\sin{\alpha_0} & \sin{\delta_0}
\end{array}\right)
\end{equation}
where ($\alpha_0$, $\delta_0$) are the coordinates of the SMC center. Then new Cartesian coordinates ($X,Y,Z$) are:
\begin{equation}
\left(\begin{array}{c}
X\\
Y\\
Z
\end{array}\right)
=T\cdot R
\end{equation}

\begin{figure*}
\centering
\mbox{\includegraphics[width=0.58\textwidth]{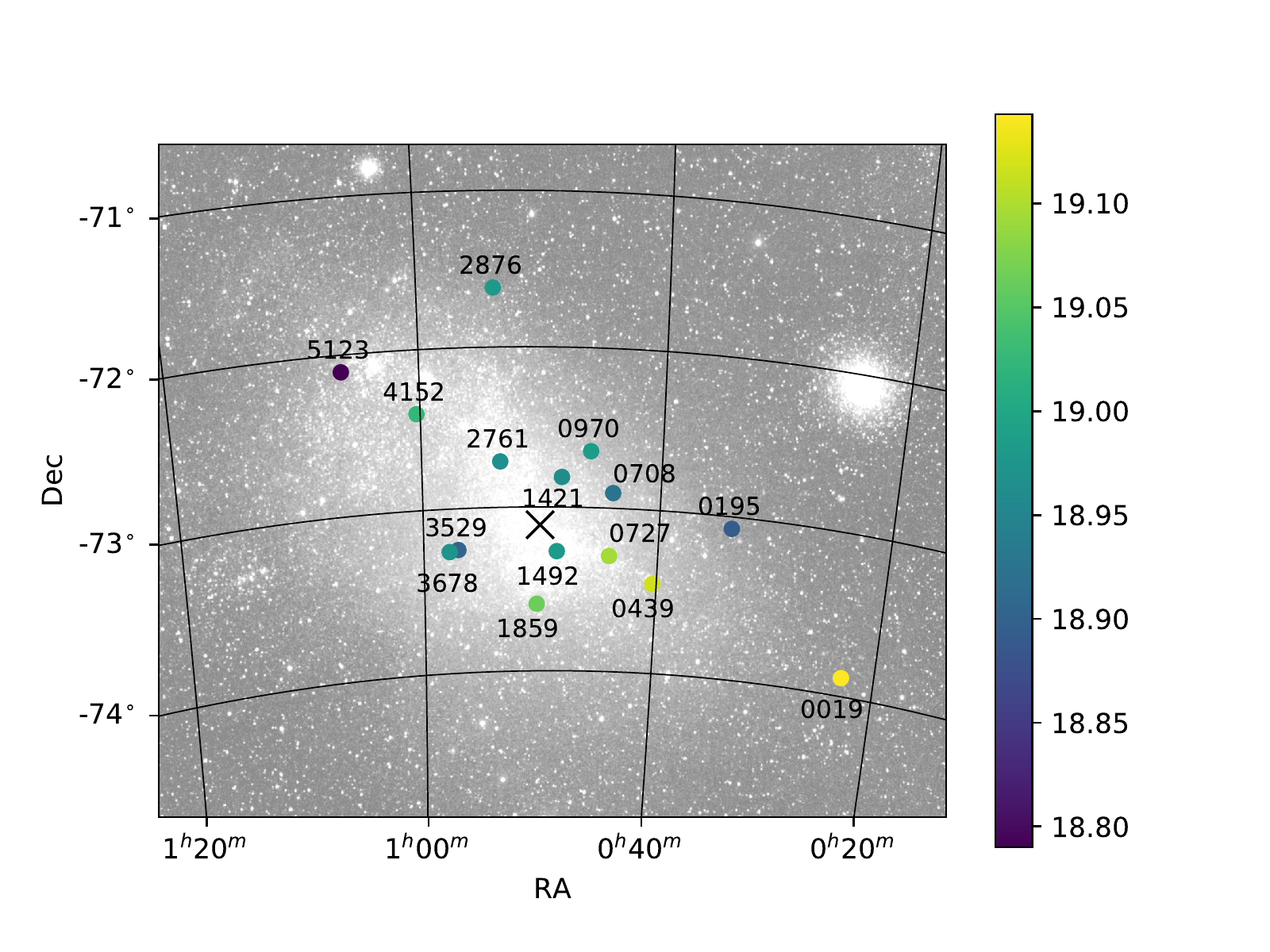}}
\hspace{-0.75 cm}
\mbox{\includegraphics[width=0.45\textwidth]{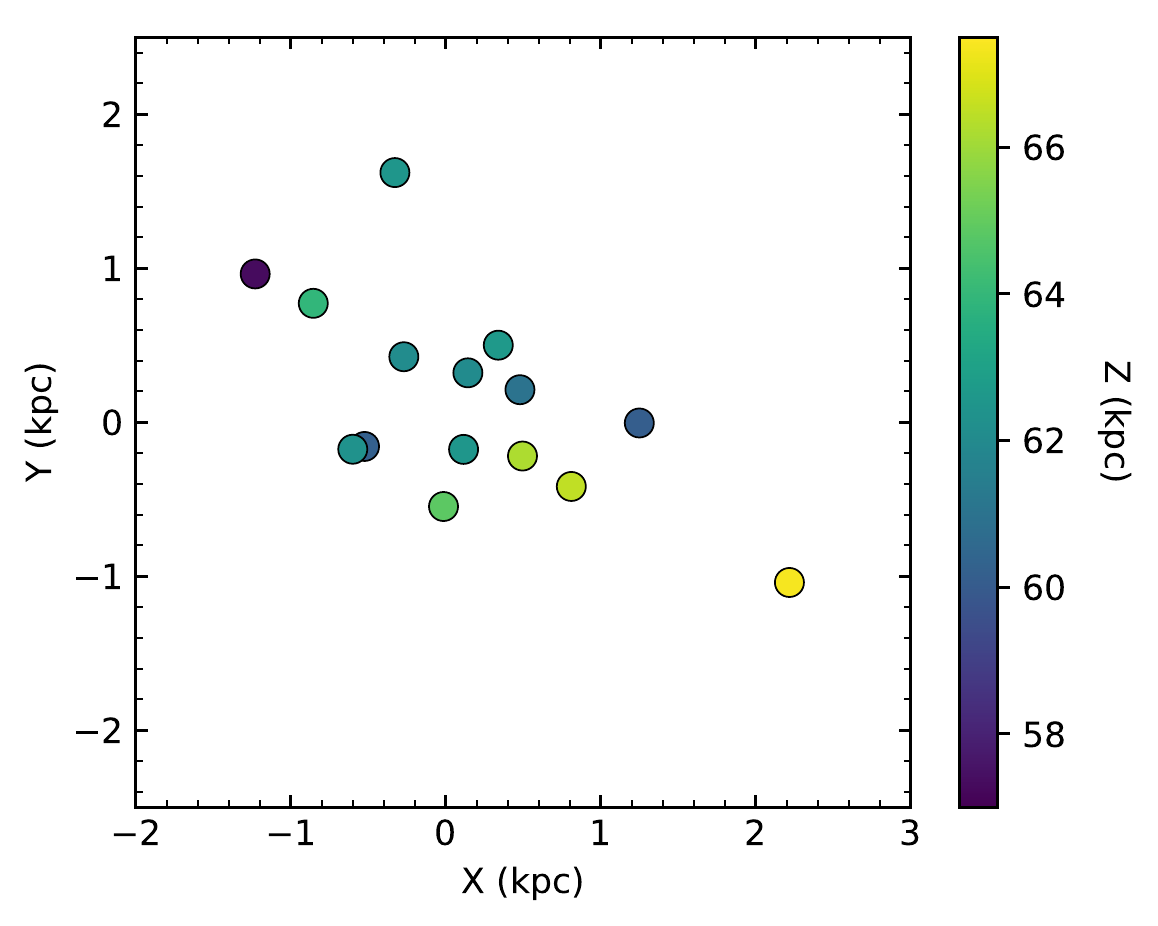}}
\caption{{\it Left:} sky positions and distance moduli of 15 eclipsing binary stars in the SMC. The cross in the middle denotes the adopted SMC center. The underlying image was obtained by ASAS and originates from \cite{uda08}. {\it Right:} the coordinates of eclipsing binaries. North is up and West is to the right. The coordinate (0,0) corresponds to the SMC center \citep{Rip17}.}
\label{fig:DXY}
\end{figure*}

Fig.~\ref{fig:DXY} shows the new coordinates of all stars in our sample. The stars show obvious distance trends in both the N-S and E-W directions. One kpc in the $XY$ plane in the Figure corresponds to about 0.9 degrees on the sky.

\subsection{Distance Determination from Analysis of Trends}
\label{trends}
\begin{figure*}
\centering
\mbox{\includegraphics[width=0.38\textwidth]{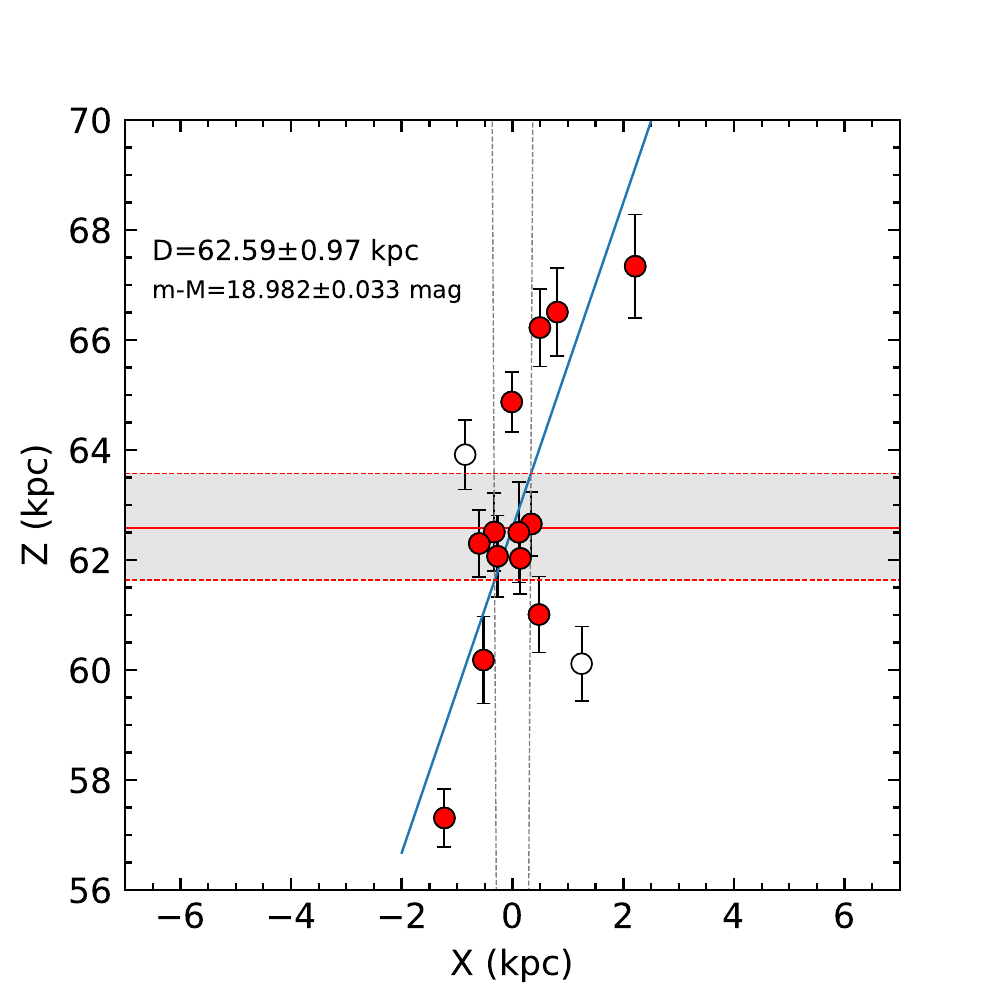}}
\mbox{\includegraphics[width=0.38\textwidth]{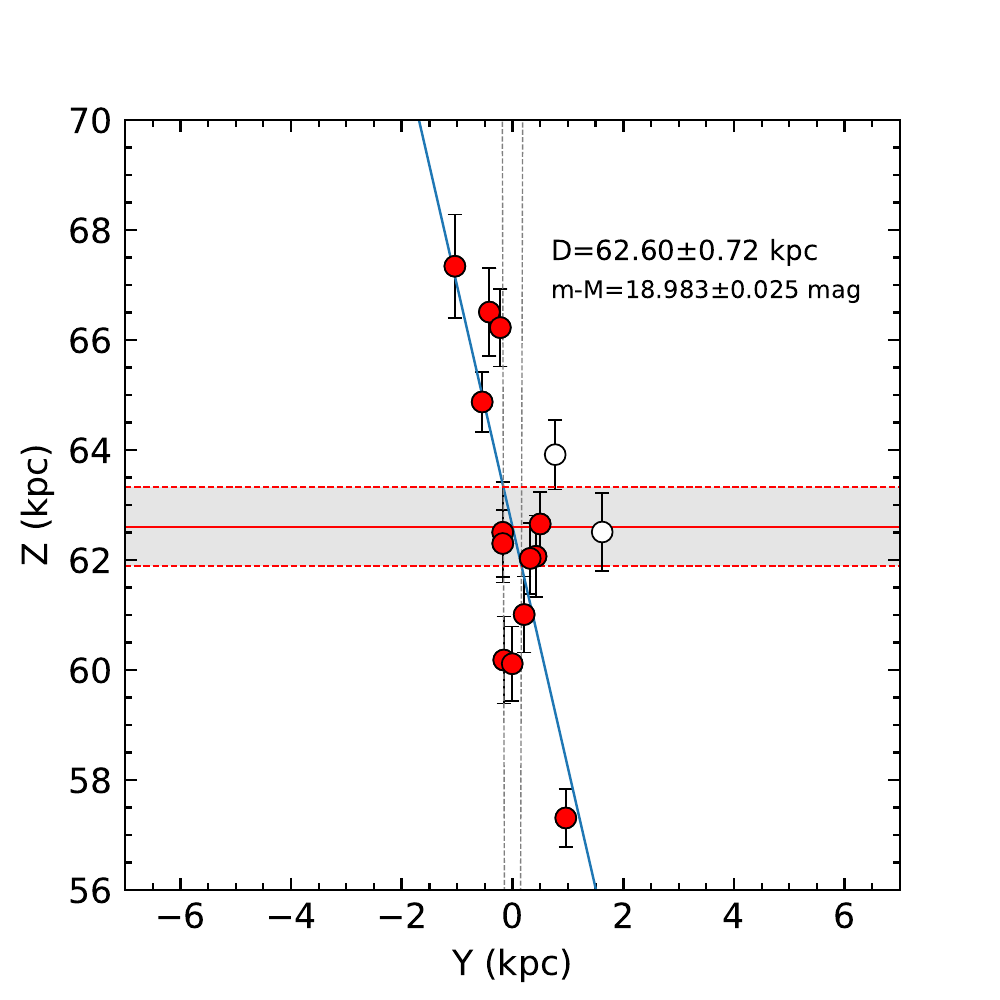}}
\caption{Position of stars on $XZ$ plane ({\it left}) and on $YZ$ plane ({\it right}). A continous blue line denotes a linear trend in distances. Vertical dashed lines denote uncertainty on position of the SMC center. Horizontal line signify a fitted distance to the SMC center and its error (shady region). Excluded outliers are denoted by empty symbol. }
\label{fig:dist_X_Y}
\end{figure*}

From a number of studies has emerged a picture of the stellar component of the SMC as an elongated,  cigar-like structure, often modeled as a triaxial ellipsoid highly inclined to the plane of the sky \citep[e.g.][]{sub12,jac16,Deb19}, with the North-East portion closer to us than the central and South-West portions. Fig.~\ref{fig:dist_X_Y} presents the positions of our targets in the $X,Z$ and $Y,Z$ planes. The sample indeed forms an elongated and highly inclined structure in agreement with previous studies: the physical extension of our sample in the $Z$ direction is 10 kpc while the tangential extensions are 3.5 kpc and 2.8 kpc in the $X$ and $Y$ directions, respectively. In both planes we fitted  linear trends $Z(X)$ and $Z(Y)$ to the positions of the stars, these are denoted as continous blue lines. We removed SMC-ECL-0195 and SMC-ECL-4152 from the sample while fitting the $Z(X)$ relation, and SMC-ECL-2876 and SMC-ECL-4152 while fitting the $Z(Y)$ trend because they are large outliers. These stars are possible members of off-axis substractures in the SMC \citep[compare with][]{jac16,Rip17}. The intersections of both trends with the '0' coordinates gives an estimate of the distance to the SMC center. The uncertainties on the  position of the SMC center are denoted as dashed, gray and almost vertical lines. The intersections of the $Z(X)$ and $Z(Y)$ trends with the uncertainty lines define the errors in the derived distance (shaded area on both panels of Fig.~\ref{fig:dist_X_Y}). The derived distance moduli of the SMC center are $m-M=18.982\pm0.033$ mag and $m-M=18.983\pm0.025$ mag from fitting $Z(X)$ and $Z(Y)$ relations, respectively. There is excellent agreement of  the distance determinations indicating the internal consistency of the method. Residuals from the fitting show that the line of sight depth of the SMC can reach up to 7 kpc.

\begin{figure*}

\centering
\mbox{\includegraphics[width=0.49\textwidth]{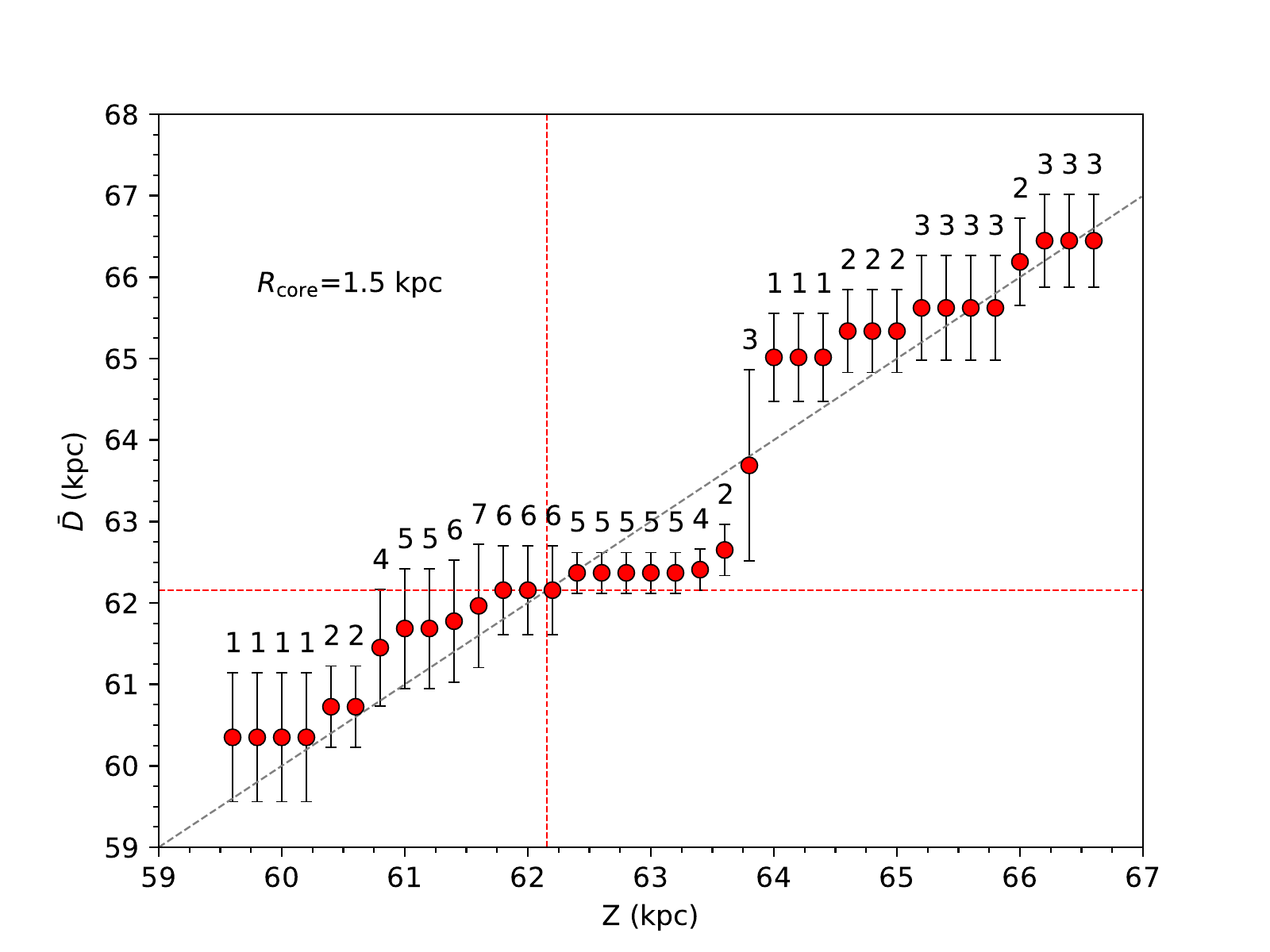}}
\mbox{\includegraphics[width=0.49\textwidth]{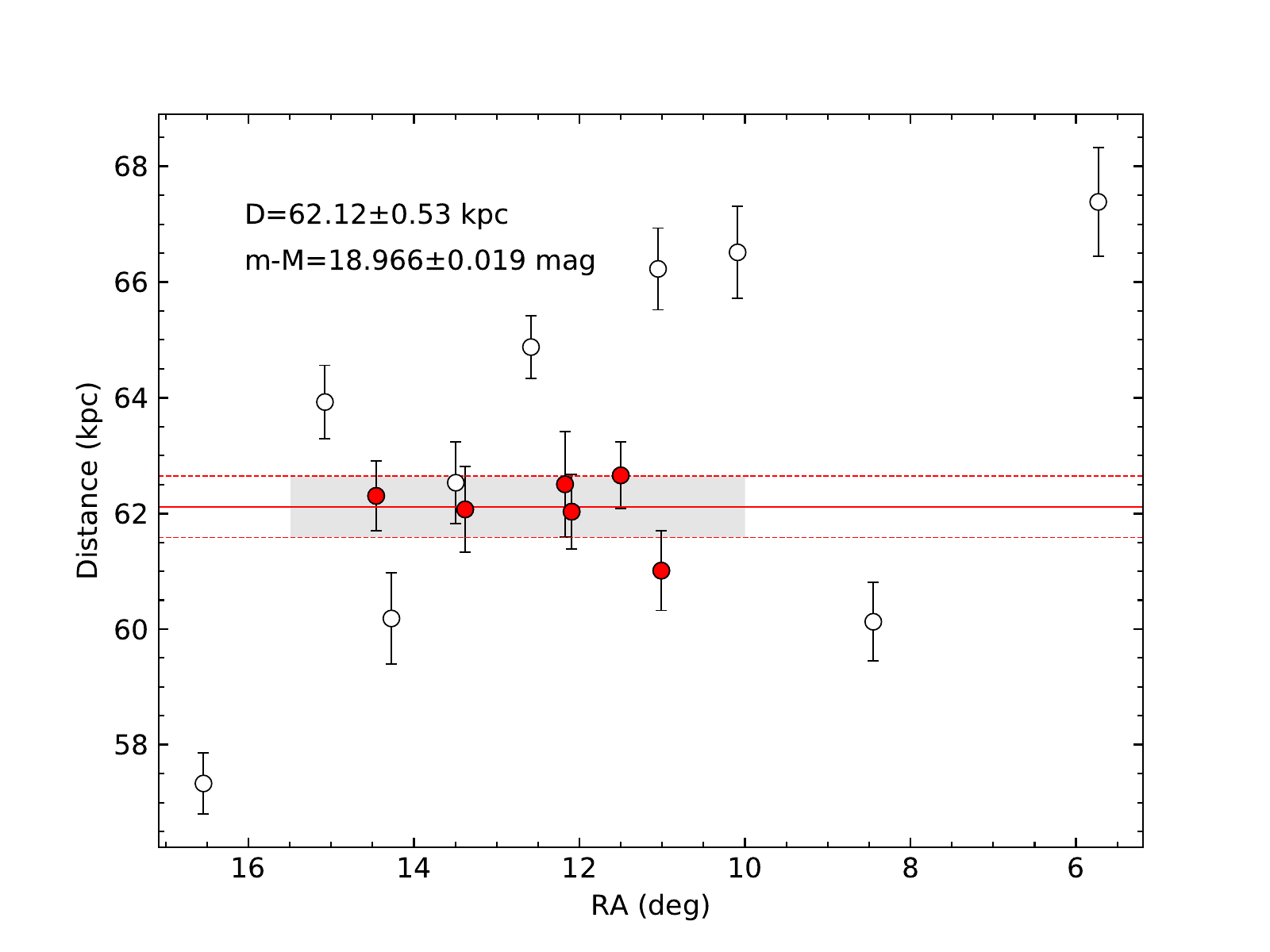}}
\caption{{\it Left:} mean weighted distance of all stars within the probe core versus the $Z$ coordinate of the probe core. Sloped dashed line denotes 1:1 relation. {\it Right:} right ascension and distances of 6 members of the central overdensity (filled symbols). Open symbols denote remaining 9 systems.}
\label{fig:cores}
\end{figure*} 

\subsection{The SMC Central Sub-structure (Core)}
\label{sec:core}
An examination of Fig.~\ref{fig:dist_X_Y} suggests that there is a central clump in the distribution of the binaries in our sample. Inspired by Sec.~5 of \cite{DeL20} we investigated the possible presence of a stellar structure with radius $<2$ kpc in the central parts of the SMC. We used a very simple approach: we moved along the trend lines $Z(X)$ and $Z(Y)$ to  probe a core of radius  $R_{core}$ at 0.2 kpc steps in the  Z axis and for each step we calculated the weighted mean of the distances $\bar{D}$ of all stars lying within the probe core. We then plotted the resulting mean distances versus the $Z$ coordinate of the core $\bar{D}(Z)$. We tested cores with radii from 1 kpc up to 2.5 kpc. For a smooth stellar distribution we would expect an almost linear function $\bar{D}(Z)$ with a slope close to 1, while the existence of substructures and overdensities would manifest as breaks in the slope of $\bar{D}(Z)$ relation and plateaus. The left panel of Fig.~\ref{fig:cores} shows an example of the $\bar{D}(Z)$ relation for $R_{\rm core}=1.5$ kpc for which the presence of the central substructure is very distinct. For a smaller radius the relation is similar (but the core is undersampled) while for a larger core radius the relation is smoother. The numbers at the top of each point indicate the number  of stars within the probe core. For core sizes near  $\sim1.5$ kpc there is a clear plateau at a distance of about 62.2 kpc. The right panel of Fig.~\ref{fig:cores} shows the 6 stars belonging to this sub-structure and their position with respect to the entire sample. The mean weighted distance modulus of the stars in the core is $m-M=18.966\pm0.019$ mag. The mean sky coordinates of the sub-structure are $\alpha=12.44$ deg, $\delta=-72.94$ deg, very close to the SMC center. Both the radial and transverse extensions of the overdensity are about $\sim1.5$ kpc.  Its position on the sky and the distance strongly suggest that this spheroidal sub-structure can be interpreted as the SMC "core".  

While we cannot rule out the possibility that the apparent core is the result of statistical fluctuation in the small number of stars
in our sample, our data are consistent with a picture of the SMC in which there exists a distinct central stellar core which is preceded and 
followed by a large number of stars (about 60\% of the total) most probably stripped by the combined tidal forces of the LMC and our Galaxy \citep[compare with e.g.][]{ziv18,DeL20}. The resulting stellar debris lies mostly in the line of sight causing the large geometrical depth of the SMC.

\subsection{Final Distance}
\label{dist-final}
The individual  distances to the SMC late-type eclipsing binaries in our sample show a large spread with a dispersion $\sigma\approx 2$ kpc (0.08 mag in distance modulus). This is much larger than the measured distances to 20 late-type eclipsing binaries in the LMC \citep[$\sigma\approx 0.6$ kpc, 0.025 mag,][]{pie19}. The distance moduli derived in Sec.~\ref{dist:ind},~\ref{trends} and~\ref{sec:core} are fully consistent with each other, and we conclude that the distance to the SMC center can be evaluated with considerable accuracy. We adopt the mean of our four distance determinations obtaining a distance of $D_{\rm SMC}=62.44$ kpc or a distance modulus $(m-M)_{\rm SMC}=18.977$ mag. We adopt the total spread of the four distance determinations as the statistical uncertainty in this estimate, a value of 0.47 kpc (0.016 mag in distance modulus). The weighted mean of our four distance determinations is only very slightly smaller: $D_{\rm SMC}=62.34$ kpc (18.974 mag), but we prefer the ordinary average because the errors are formal and not based on  modeling. Summary of all distance determination methods and final results are given in Table~\ref{tbl-dist}.

\begin{deluxetable}{@{}lccc@{}}
\tabletypesize{\scriptsize}
\tablecaption{Distance determination to the SMC \label{tbl-dist}}
\tablewidth{0pt}
\tablehead{\colhead{Method} &\colhead{Distance} &\colhead{Error} &\colhead{Reference}\\
\colhead{}&\colhead{kpc}&\colhead{kpc}&\colhead{}
}
\startdata
 Weighted mean from all &  62.45 & 2.61 & Sec.~\ref{dist:ind} \\
 Distance trend in X axis &  62.59 & 0.97 & Sec.~\ref{trends} \\
 Distance trend in Y axis & 62.60 & 0.72 & Sec.~\ref{trends} \\
 Core distance                &  62.12 & 0.53 &  Sec.~\ref{sec:core} \\\hline
 Weighted mean of above & 62.34 & 0.24 & Sec.~\ref{dist-final} \\
 {\bf Unweighted mean of above }& {\bf 62.44} & {\bf 0.47} & {\bf Sec.~\ref{dist-final}} 
\enddata
\end{deluxetable}

\subsection{Systematic Uncertainty}
\label{sec:sys}
The systematic error combines uncertainty in the SBC calibration, uncertainty in the zero-point of the optical $V$-band photometry, and uncertainty on the extinction law \citep{pie19}.
To quantify the systematic error of the zero-point of $K$-band photometry we included additional NIR photometry from VMC, IRSF, and 2MASS, (Sec.~\ref{obs:phot}) and we repeated all steps of the distance determination. We ended up with similar conclusions (clearly visible N-S, E-W distance trends, existence of the central overdensity) and the resulting  distance modulus is larger by only 0.008 mag. We assigned an additional zero-point error in the $K$-band photometry  of 0.01 mag to the total error budget.

A recent paper by \cite{mou19} based on an analysis of theoretical spectra suggests the existence of non-negligible metallicity corrections to the surface brightness $(V\!-\!K)$ - color relation for the low metallicity range of both Magellanic Clouds. However our empirical data show that the sensitivity of the SBC calibration on metallicity is completely negligible to within a precision of 1\% \citep[see Extended Data Fig.\,2 in][]{pie19}. In accordance with it we do not apply any metallicity correction to the SBC relation by Pietrzy{\'n}ski et al.

The OGLE project has published extensive reddening maps of the Magellanic Clouds derived from observations of red clump stars \citep{sko20}. The intrinsic colors of the red clump stars were calibrated  using fields distant from the centers of both galaxies where only Milky Way foreground extinction is expected.  A comparison with reddening estimates from the OGLE maps shows that the values reported in Table~\ref{tbl-5} are larger by 0.020 mag on average. Such a systematic shift would result in the distance modulus of the SMC becoming larger by 0.013 mag (0.6\%). In \cite{pie19} we assumed a zero-point error of the calculated reddenings of 0.02 mag. Here we use the same approach to derive the interstellar extinction, and we conclude that to within this uncertainty our reddenings are consistent with the zero-point of the recent OGLE maps. The resulting systematic error of 0.6\% is already in the total error budget. Finally, we adopt a total systematic error of 0.028 mag or 0.81 kpc.

\section{Concluding Remarks}
\label{final}
By combining the distances of 15 eclipsing binary stars we have obtained a distance modulus to the SMC of $\mu_{\rm SMC}=18.977\pm 0.016\pm 0.028$ mag. This significantly improves on the precision and the accuracy of the distance determination to the SMC from our previous report of $\mu_{\rm SMC}=18.965\pm0.025\pm0.048$ mag (G14). The improvement results from the use of a more accurate surface brightness - color relation  for giant stars \citep{pie19} and by a 3-fold increase in the sample of binary stars. We have used exactly the same observational setup and reduction methodology as for case of the LMC, and  the resulting distance modulus difference between these galaxies is very well determined: $\Delta\mu=\mu_{\rm SMC}-\mu_{\rm LMC}=0.500\pm0.017$ mag. Comparison with Fig.~9 from G14 shows that our value of $\Delta\mu$ is close to the peak of $\Delta\mu$ distribution for a number of distance indicators.

An extensive set of recent distance determinations to the SMC based on a number of independent methods was analysed by \cite{deG15}. They advocated a distance modulus to the SMC  of $18.96\pm0.02$ mag which is in perfect agreement with our result. The distance we report here is also in full accordance with most recent distance determinations based on Classical Cepheids \citep[e.g.][]{sco16,rip16}.   

This is our final paper regarding the distance determination to the SMC with the eclipsing binary method. We started spectroscopic follow-up of the eclipsing binary OGLE SMC-ECL-5123  in 2002, and thus we report here results from an almost 18-year long observational effort. 

\facilities{ ESO 3.6-m, Magellan Clay 6.5-m, ESO NTT, and VLT 8-m telescopes}

\acknowledgments
We are deeply indebted to the OGLE team, especially to prof. A. Udalski and prof. I. Soszy{\'n}ski
for making available to us early versions of the eclipsing binary catalogues in the SMC.

The research leading to these results  has received
funding from the European Research Council (ERC) under the European
Union's Horizon 2020 research and innovation program (grant agreement
No 695099) and from the National Science Center, Poland grants MAESTRO 
UMO-2017/26/A/ST9/00446 and BEETHOVEN  UMO-2018/31/G/ST9/03050.
We acknowledge support from the IdP II 2015 0002 64 and DIR/WK/2018/09 
grants of the Polish Ministry of Science and Higher Education. \newline
Support from the BASAL Centro de Astrof{\'i}sica y Tecnolog{\'i}as Afines (CATA) AFB-170002, the
Millenium Institute of Astrophysics (MAS) of the Iniciativa Cientifica
Milenio del Ministerio de Economia, Fomento y Turismo de Chile, project
IC120009.

M.T. acknowledges financial support from the Polish National Science
Centre grant PRELUDIUM 2016/21/N/ST9/03310.

R.P.K. has been supported by the Munich Excellence cluster ORIGINS funded by the Deutsche Forschungsgemeinschaft (DFG, German Research Foundation) under Germany's Excellence Strategy EXC-2094-390783311. 

Based on observations made with ESO 3.6m and NTT telescopes in La Silla under programme 074.D-0318, 074.D-0505, 082.D- 0499, 083.D-0549, 084.D-0591, 086.D-0078, 091,D-0469(A), 0100.D-0339(A), 098.D-0263(A,B), 097.D-0400(A), 0102.D-0469(B) and 0102.D-0590(B). We also acknowledge the generous allocation of CNTAC time and time allocated by the Carnegie Observatories TAC.
 
{}    

\listofchanges
\end{document}